\documentclass[aps,reprint,superscriptaddress,twocolumn,longbibliography ]{revtex4-1}

\usepackage{amsmath,amsthm,amssymb}

\usepackage{graphicx}
\usepackage{dcolumn}
\usepackage{bm}

\usepackage[utf8]{inputenc}
\usepackage{mathptmx}
\usepackage{upgreek}
\usepackage{xfrac}

\usepackage{epstopdf}

\begin{document} 
	%
	%
	%
	%
	%
	\title{How Nanoporous Silicon-Polypyrrole Hybrids\\ Flex Their Muscles in Aqueous Electrolytes: In Operando High-Resolution X-Ray Diffraction and Electron Tomography-Based Micromechanical Computer Simulations}
	%
	%
	\author{Manuel Brinker}
	\altaffiliation{manuel.brinker@tuhh.de}
	\author{Marc Thelen}
	\author{Manfred May}
	\affiliation{Hamburg University of Technology, Institute for Materials and X-Ray Physics, Denickestr. 10, 21073 Hamburg, Germany}
	\affiliation{Center for X-Ray and Nano Science CXNS, Deutsches Elektronen-Synchrotron DESY, Notkestr. 85, 22607 Hamburg, Germany}
	\affiliation{Center for Hybrid Nanostructures CHyN, Hamburg University, Luruper Chaussee 149, 22607 Hamburg, Germany}

	\author{Dagmar Rings}
	\affiliation{Institute of Materials Physics and Technology, Hamburg University of Technology, 21073 Hamburg, Germany}
	\author{Tobias Krekeler}
	\affiliation{Electron Microscopy Unit BEEM, Hamburg University of Technology, 21073 Hamburg, Germany}
	\author{Pirmin Lakner}
	\author{Thomas F. Keller}
	\affiliation{Physics Department, University of Hamburg, 20355 Hamburg, Germany}
	\affiliation{Center for X-Ray and Nano Science CXNS, Deutsches Elektronen-Synchrotron DESY, Notkestr. 85, 22607 Hamburg, Germany}
	\author{Florian Bertram}
	\affiliation{Deutsches Elektronen-Synchrotron DESY, Notkestr. 85, 22607 Hamburg, Germany}
	\author{Norbert Huber}
	\affiliation{Institute of Materials Research, Materials Mechanics, Helmholtz-Zentrum Hereon, 21502 Geesthacht, Germany}
	\affiliation{Institute of Materials Physics and Technology, Hamburg University of Technology, 21073 Hamburg, Germany}
	\author{Patrick Huber}
	\altaffiliation{patrick.huber@tuhh.de }
	\affiliation{Hamburg University of Technology, Institute for Materials and X-Ray Physics, Denickestr. 10, 21073 Hamburg, Germany}
	\affiliation{Center for X-Ray and Nano Science CXNS, Deutsches Elektronen-Synchrotron DESY, Notkestr. 85, 22607 Hamburg, Germany}
	\affiliation{Center for Hybrid Nanostructures CHyN, Hamburg University, Luruper Chaussee 149, 22607 Hamburg, Germany}
	\begin{abstract}	
		Macroscopic strain experiments have revealed that silicon crystals traversed by parallel, channel-like nanopores functionalized with the artificial muscle polymer polypyrrole (PPy) exhibit large and reversible electrochemo-mechanical actuation in aqueous electrolytes. On a macroscopic scale these actuation properties are well understood. However, on the microscopical level this system still bears open questions, as to how the electrochemical expansion and contraction of PPy acts on to np-Si pore walls and how the collective motorics of the pore array emerges from the single-nanopore behavior. Here we present synchrotron-based, in operando X-ray diffraction on the evolving electrostrains in epilayers of this material grown on bulk silicon. An analysis of these experiments with micromechanical finite element simulations, that are based on a full 3D reconstruction of the nanoporous medium by transmission electron microscopy (TEM) tomography, shows that the in-plane mechanical response is dominantly isotropic despite the anisotropic elasticity of the single crystalline host matrix. However, the structural anisotropy originating from the parallel alignment of the nanopores lead to significant differences between the in- and out-of-plane electromechanical response. This response is not describable by a simple 2D arrangement of parallel cylindrical channels. Rather, the simulations highlight that the dendritic shape of the silicon pore walls, including pore connections between the main channels, cause complex, highly inhomogeneous stress-strain fields in the crystalline host. Time-dependent X-ray scattering experiments on the dynamics of the actuator properties hint towards the importance of diffusion limitations, plastic deformation and creep in the nanoconfined polymer upon (counter-)ion adsorption and desorption, the very pore-scale processes causing the macroscopic electroactuation. From a more general perspective, our study demonstrates that the combination of TEM tomography-based micromechanical modeling with high-resolution X-ray scattering experiments provides a powerful approach for in operando analysis of nanoporous composites from the single-nanopore up to the porous-medium scale.    
	\end{abstract}	
	\maketitle
	\section*{Introduction}
	Since the discovery of self-organized nanoporosity in silicon during electrochemical etching the resulting material has been attracting much attention from both fundamental and applied sciences owing to its exceptional biocompatible~\cite{Park2009,Chiappini2015}, optical, electrical and thermal properties compared to the bulk material~\cite{Canham1990,Lehmann1991,Sailor2011,Nattestad2010,Canham2015,Henstock2015,Gor2015,Huber2015}. Nanoporous silicon (np-Si) can be prepared in monolithic form with parallel, channel-like pores and has been of particular interest for studies on the influence of geometrical confinement on condensed-matter systems~\cite{Gruener2008, Henschel2009,Kusmin2010,Hofmann2013,Kondrashova2017, Bossert2021}. It also allows one to synthesize biocompatible composite materials with finely tuned optical and electrical properties by liquid adsorption or infiltration~\cite{Westover2014,Canham2015,Brinker2021}. Moreover, by thermal oxidation in can be transformed pseudomorphically to nanoporous silica glass that can be used as model geometry for studies of spatially nanoconfined matter~\cite{Huber2015,Kityk2008,Sentker2018}.\\
	Here, we combine this 3D scaffold structure with the semiconducting polymer polypyrrole (PPy). PPy has attracted increasing interest because of its biocompatibility~\cite{Skotheim2007} and its tunable conductivity, that can be as high as carbon's~\cite{Satoh1986}. The electrical conductivity has its origin in delocalized charge carriers located along the polymer backbones. An electrochemical change in the oxidation state of the polymer can vary the number of delocalized charges~\cite{Madden2002}. Apart from the electrical conductivity, a change in oxidation state of the polymer has an impact on the optical properties~\cite{Bredas1985} of the polymer, the ionic permeability~\cite{Burgmayer1982} and the mechanical properties~\cite{Otero2007,Shoa2010b}. The change in oxidation state is accompanied by a balancing of the respective ions solved in a surrounding electrolyte which maintains the charge neutrality of PPy~\cite{Madden2002}. The intercalated ions cause the polymer to swell~\cite{Pei1993}, which makes PPy an interesting material for actuoric applications~\cite{Madden2000}. Its stability upon electrochemical cycling~\cite{Shoa2010a} and a low operating voltage~\cite{Madden2000} are adding to this. These characteristics have led to an increased investigation into biomedical devices~\cite{Smela2003}, microactuators~\cite{Jager2000}, biomimetic robots~\cite{McGovern2010} and artificial muscles~\cite{Otero1996,Madden2004}. In all these processes the actuation depends on the transport of ions into PPy.\\
	Thin-film geometries are suitable to counter a possible retardation of the actuation process by improving the kinetics of the ion exchange~\cite{Mirfakhrai2007}. PPy films in planar geometry also permit to determine the anion induced polymer swelling by in situ X-ray reflectivity (XRR), which provides insight into the ion incorporation mechanisms, due to its inherent sensitivity to electron densities and an additional comparison of electrochemical data. Thereby, the overall volume change of the PPy film might include charge compensation by protons and a counterflow of solvent molecules~\cite{Lakner2020}. Yet, these thin-film properties are counterbalanced by possible restrictions through creep~\cite{Spinks2006,Smela2005} and a low sustainable stress during actuation~\cite{Sonoda2001,Spinks2002}. Furthermore, the thin-film geometry needs a supporting substrate on which the PPy film is synthesized, such as bilayered bending cantilevers~\cite{Smela1995} or PPy coated conductive fibers~\cite{Hutchison2000}. Thus, these thin-film PPy actuating structures are heavily dependent on their substrate.\\	
	A strategy to enhance the stiffness and strength is to use np-Si as a mechanically rigid and robust scaffold material as some of us recently reported for a nanoporous silicon polypyrrole (np-Si/PPy) hybrid~\cite{Brinker2020}. This composite material shows a macroscopic electrochemical actuation effect that is controllable by an applied potential in a highly linear and reversible fashion. Therefore, a direct electromechanical control of this mainstream semiconductor is achieved. In a macroscopic sense the electrochemical actuation properties of this hybrid system are well investigated~\cite{Brinker2020}. On the microscopical level this system still bears open questions, as to how the electrochemical expansion and contraction of PPy acts on to np-Si pore walls. Therefore, we present here a comprehensive in-situ X-ray diffraction (XRD) study on PPy swelling induced straining of the lattice in np-Si pore walls.\\
	An XRD study conducted on a np-Si layer attached to an underlying bulk silicon substrate constitutes an elegant method to scrutinize with high resolution lattice strains in np-Si, also under in-situ conditions like adsorption or desorption of liquids~\cite{Barla1984,Bellet1994,Bellet1996,Buttard1996,Dolino1996}. To probe a changing lattice strain in the np-Si layer, $\theta-2\theta$ scans of a Bragg peak can be conducted. The strain in the np-Si lattice will result in a change of the lattice constant with respect to the underlying bulk silicon. Thus, distinct Bragg peaks for both, the porous- and the bulk silicon layer, can be measured. The lattice parameters of the np-Si layer can then be determined with respect to bulk silicon's by the splitting between the peaks. The measurement method does not depend on the actual peak position of both Bragg peaks but only on the relative position and can therefore be considered independent of measurement parameters, e.g., a perfect alignment. Moreover, the underlying bulk silicon has approximately an order of magnitude larger thickness than the thickness of the np-Si layer. Thus, it acts as a stabilization and is supposed to prevent a bending of the bi-layered sample which would have an impact on the actuation and could distort also the XRD signal.\\ 
	To connect the silicon lattice electrostrain experiments with the single-nanopore and porous-medium mechanical response we will employ micromechanical simulations, that are based on a transmission electron microscopy (TEM) tomography of the nanoporous host. This approach provides a deep look into the microscopic strain distribution and links it to the macroscopic strain. \\
	\section*{Materials and Methods}
	\subsection{Porous silicon fabrication and electropolymerization in pore space }
	A np-Si/PPy layer attached to a thicker bulk silicon support underneath is synthesized as follows. In a first step, pores are etched electrochemically into a bulk silicon wafer. Subsequently, the pores are filled by the electrically conductive polymer PPy, which is responsible for the electrochemical actuation in the np-Si pores. PPy is prepared directly within the pore space by an electrochemical electropolymerization procedure.\\
	The exact fabrication route has been published before~\cite{Brinker2020}. Thus, the sample synthesis will be described only in a brief overview. The np-Si layer is prepared in an electrochemical etching procedure with hydrofluoric acid (HF) as the electrolyte solution. The wafer is installed into an electrochemical etching cell. The area that is in contact with the electrolyte solution in the cell is $7.84\,\mathrm{cm^2}$. The wafer is electrically contacted as the working electrode (WE) and a platinum counter-electrode (CE) is inserted into the cell. To etch the pores into silicon, a current of $98\,\mathrm{mA}$ is applied between the CE and the silicon wafer for 26 minutes and 40 seconds. This results in a np-Si layer thickness of $t_\mathrm{np-Si}=25.4\,\mathrm{\upmu m}$, a value determined by a scanning electron micrograph profile. The resulting porous layer has a porosity of $\Phi=45.7\,\%$. This value is determined by an analysis of nitrogen sorption isotherm. The respective measurement is displayed in Figure S1 in the supplementary materials\cite{Supp2022}.\\
	Afterwards the pores are filled with PPy. The sample is installed into a second electrochemical cell. In addition to a platinum CE, a silver-silverchloride (Ag/AgCl, Sensortechnik Meinsberg) reference electrode (RE) is present in the cell. The electropolymerisation solution consists of the monomeric unit of the polymer, a $\mathrm{C_4H_4NH}$ molecule with a concentration of $0.1\,\mathrm{mol\,l^{-1}}$. The resulting PPy polymer is not in a neutral state, but is positively charged~\cite{Madden2002}. To counter balance these charges, lithium perchlorate ($\mathrm{LiClO_4}$) electrolyte with a concentration of  $0.1\,\mathrm{mol\, l^{-1}}$ is added to the electropolymerisation solution. Acetonitrile is used as a solvent. The polymerisation itself is carried out by applying a current of $2\,\mathrm{mA}$. Specific phases of the polymerization process can be monitored through the potential evolution recorded by the RE in steps of $1\,\mathrm{s}$, as shown in Figure S2 in the supplementary materials\cite{Supp2022} along with further information on the process.
	\subsection{X-ray diffraction measurements}
	Two samples $\mathrm{S_{refl}}$ and $\mathrm{S_{trans}}$ are prepared from the same PPy infiltrated np-Si material by cleaving~\cite{Sailor2011}. The samples are bar shaped and have a width of approximately $3\,\mathrm{mm}$ and a length $20\,\mathrm{mm}$. The two samples are designated for two types of X-Ray diffraction (XRD) measurements. The samples are placed onto a PTFE sample holder and electrically contacted on the backside by aluminum foil and a gold wire. The contact is then electrically insulated by a thermoplastic film. These sample holders can then be mounted into the in-situ measuring cell. The cell enables electrochemical control and simultaneous XRD measurements. The cell is filled with electrolyte solution with a defined volume of $7.1\,\mathrm{mL}$. The electrolyte solution is perchloric acid ($\mathrm{HClO_4}$) with a concentration of $1\,\mathrm{mol\,l^{-1}}$. The sample is then installed in an upright hanging way, so that it is immersed in the reservoir of electrolyte solution. Thereby, the surface area that is in contact with the acid amounts to $A_\mathrm{refl}=46.61\,\mathrm{mm^2}$ and $A_\mathrm{trans}=63.185\,\mathrm{mm^2}$. These values are of importance, when determining the sample specific electrochemical capacitance. An Ag/AgCl-RE and a carbon-cloth-CE are installed and inserted into the acid. The sample is connected to a potentiostat (Metrohm-Autolab PGSTAT 30).\\
	\begin{figure}
		\centering
		\includegraphics{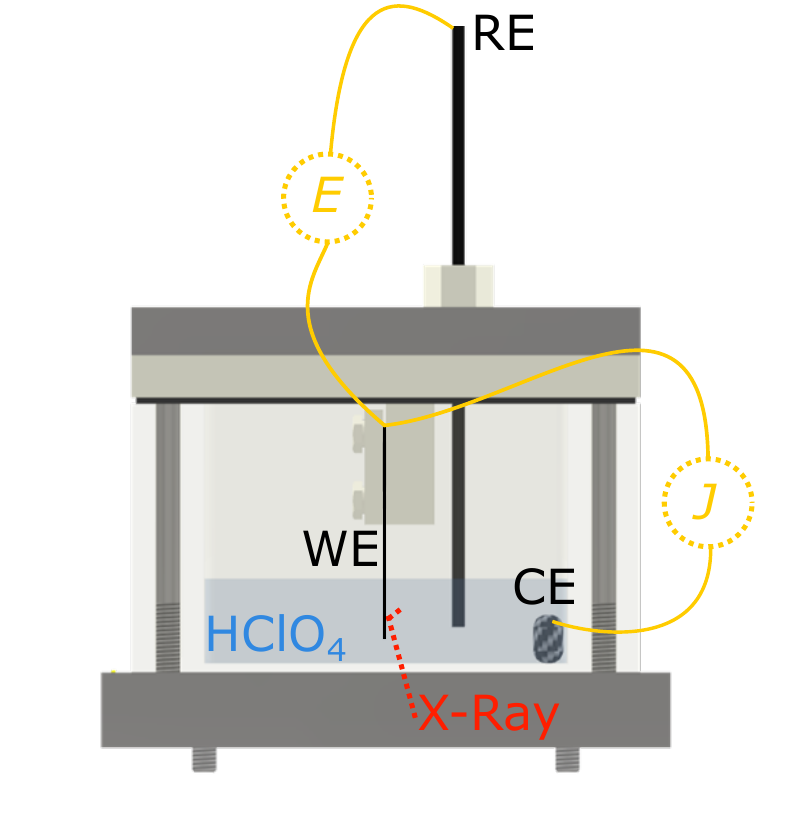}
		\caption{\textbf{In-situ electrochemical measurement cell for X-ray diffraction.} In-situ XRD measurement cell with the sample, acting as the working electrode, in the center, the reference electrode, a counter electrode and the electrolyte solution. Depicted are as well the measurement sites for potential $E$ and current $J$ and the probing X-Ray beam. In the area where the X-Ray beam penetrates the in-situ measuring cell the strength of the cell wall is significantly reduced to $2\,\mathrm{mm}$, so that the diffracted beam has still a high enough intensity when leaving the cell.}
		\label{Fig_Cell}
	\end{figure}
	The X-Ray diffraction measurements are conducted at beamline P08 of the $3^{\mathrm{rd}}$ generation synchrotron PETRA III at Deutsches Elektronen-Synchrotron (DESY). The cell is installed onto a 6-circle closed Eulerian cradle which enables a comprehensive covering of possible diffraction geometries. The energy of the photon beam is adjusted to $E=18\,\mathrm{keV}$ with a spot size of $200\times200\,\mathrm{\upmu m^2}$. A sketch of the cell and the beam penetrating the cell is given in Figure \ref{Fig_Cell}. The diffracted signal is measured by a 2D detector (Pilatus 100K, pixel number $487 \times 195$ (horizontal $\times$ vertical), pixel size $p$ $172\times172\,\mathrm{\upmu m^2}$). The signal itself is evaluated by summing the intensity of a region in the central detector plane. The distance from the sample to the detector is determined as $D=(955.6 \pm 1.4)\,\mathrm{mm}$, which is discussed in the supplementary methods in Figure S3\cite{Supp2022}.\\
	The measurements are conducted as $\theta$-$2\theta$ scans. Hence, crystal planes with the interplanar distance $d$ are probed and $d$ can then be calculated according to Bragg's law
	\begin{equation}
		\label{Eq_Bragg}
		\lambda =2 d\sin(\theta),
	\end{equation}
	where $\lambda=\mathrm{c\cdot h}\,E^{-1}$ with c and h denoting speed of light and Planck's constant. The lattice constant $a$ of the material can then be obtained by $d$ and the respective $hkl$-index of the probed planes. Note, that, as discussed in the introduction, the lattice constant of the bulk silicon layer and the np-Si pore walls within the adjacent layer is different. The lattice constant of np-Si, $a_\mathrm{np-Si}$, can then be directly related to the difference of the measured angles of both bulk- and np-Si, i.e. $\theta_\mathrm{Si}$ and $\theta_\mathrm{np-Si}$ respectively, through
	\begin{equation}
		\label{Eq_da/a}
		\frac{a_\mathrm{np-Si}-a_\mathrm{Si}}{a_\mathrm{Si}}=\frac{\delta a}{a_\mathrm{Si}}=\frac{\delta \theta}{\tan(\theta_\mathrm{Si})}=\frac{\theta_\mathrm{Si}-\theta_\mathrm{np-Si}}{\tan(\theta_\mathrm{Si})},
	\end{equation}
	where $a_\mathrm{Si}=0.543\,\mathrm{nm}$ denotes the lattice constant of bulk silicon~\cite{Dolino1996,Lehmann2002}. Hence, a positive value for $\delta a/a_\mathrm{Si}$ corresponds to a larger np-Si lattice constant with respect to the one of bulk silicon and thus means an expanded np-Si lattice, and vice versa. Equation \ref{Eq_da/a} does not require a determination of absolute values for $\theta$ neither for the porous- nor the bulk silicon layer. Rather, the relative value of $\delta \theta$ determines the change in lattice constant. The change in lattice constant $\delta a/ a_\mathrm{Si}$ is here referred to as the lattice mismatch and abbreviated as $\delta a/a_\perp$ for the out-of-plane direction and $\delta a/a_{||}$ for the in-plane direction. Please note, that Equation \ref{Eq_da/a} warrants values for $\theta$ in radians.\\ 
	The angular uncertainty $\delta\theta$ of the X-ray diffraction measured at the detector can be estimated by\cite{Chernyshov2021}
	\begin{multline}
		\label{Eq_error}
		(\delta\theta)^2=\frac{\cos^4(2\theta)}{16D^2}\left[\tan^2(2\theta)(t^2 + 2h^2)+p^2\right.\\+\frac{t^2}{\cos^2(2\theta)}+ \left.\frac{D^2\varphi^2}{\cos^4(2\theta)}\right],
	\end{multline}
	where $D$ denotes the detector distance and $t$ the sample thickness, which is set to the beam size in the case of the reflection, out-of-plane measurement. $h$ is the thickness of the detection layer, $p$ the pixel size and $\varphi=12\,\mu \mathrm{rad}$ is the vertical beam divergence. Further, for the uncertainties of the fit of a $\theta-2\theta$ diffraction measurement, the number of measurement points needs to be taken into account. For visualization, Figure S4 in the supplementary materials\cite{Supp2022} shows the silicon diffraction peak in an out-of-plane measurement with the respective error bars for each measurement point. For the lattice mismatch, the uncertainty is obtained by a propagation of uncertainty of Equation \ref{Eq_da/a}. The uncertainties given here for the lattice mismatches originate from these considerations. Exceptions are the amplitudes $A_{\delta a/a_\perp}$ and $A_{\delta a/a_{||}}$ from the constant potential scans, see Figures \ref{Fig_XRD-out-ofplane}(e) and \ref{Fig_XRD_in-plane}(d). Here, the uncertainty is determined by the averaging of the single lattice mismatch values over several of the potential cycles.\\
	Two types of samples are used for two types of XRD measurements. Thereby, the two samples are probed under different conditions. For one, planes with a $hkl$-index of (400) are measured under an angle of $2\theta_\mathrm{refl}=29.3934\,^{\circ}$, according to Equation \ref{Eq_Bragg}. Thus, the incident beam hits the surface of the np-Si layer under an angle of $\theta_\mathrm{refl}$ and the detector is positioned with an equal angle of $\theta_\mathrm{refl}$. Hence, planes that run parallel to the surface are probed and Equation \ref{Eq_da/a} yields the change in lattice constant orthogonal to the surface $\delta a/a_\perp$. For the other measurement type, the incident beam hits the sample on its backside so that it first passes through the bulk silicon- and then the np-Si layer. To probe the (440) lattice planes that are orthogonal to the surface an angle of $\theta_\mathrm{trans}=42.0517\,\mathrm{^{\circ}}$ is chosen in accordance to Equation \ref{Eq_Bragg}. Thus, the in-plane difference in lattice constant $\delta a/a_{||}$ is determined. A detailed depiction of the two measurement geometries are presented in Figure \ref{Fig_XRD-out-ofplane}(a) and Figure \ref{Fig_XRD_in-plane}(a).\\
	\subsection{Electrochemical actuation}
	The potentiostat connected to the electrodes of the measurement cell can apply a potential $E$ between the sample and the CE controlled via the RE. The potential values in this work are indicated versus the standard hydrogen electrode potential (SHE). Simultaneously, the current $j$ that flows between sample and CE is measured and hence the consumed charge $Q$ as well. For a so called cyclic-voltammetry (CV) measurement, the potential $E$ is linearly increased from a lower- to an upper vertex point and back, repeatedly for several cycles. The rate $\mathrm{d}E/\mathrm{d}t$ of the potential change is called the scan rate. Through these CV measurements details about the electrochemical characteristics can be inferred. The anions of the solution can be incorporated into and extracted from the polymer by means of applying a potential to the sample. The polymer accordingly expands and contracts upon the anion movement. Furthermore it is possible to asses, whether the sample possesses a capacitive potential region. Thereby, increasing $E$ from its lower vertex point leads to an almost instant increase in $j$.
	The magnitude of the current and the scan rate then determine the capacity $c$ by~\cite{Roschning2019}
	\begin{equation}
		\label{Eq:capacity}	
		c=j\cdot (\mathrm{d}E/\mathrm{d}t)^{-1}.
	\end{equation}
	Here a potential window of $0.4$ to $0.8\,\mathrm{V}$ is chosen. 
	\begin{figure*}[]
		\centering
		\includegraphics{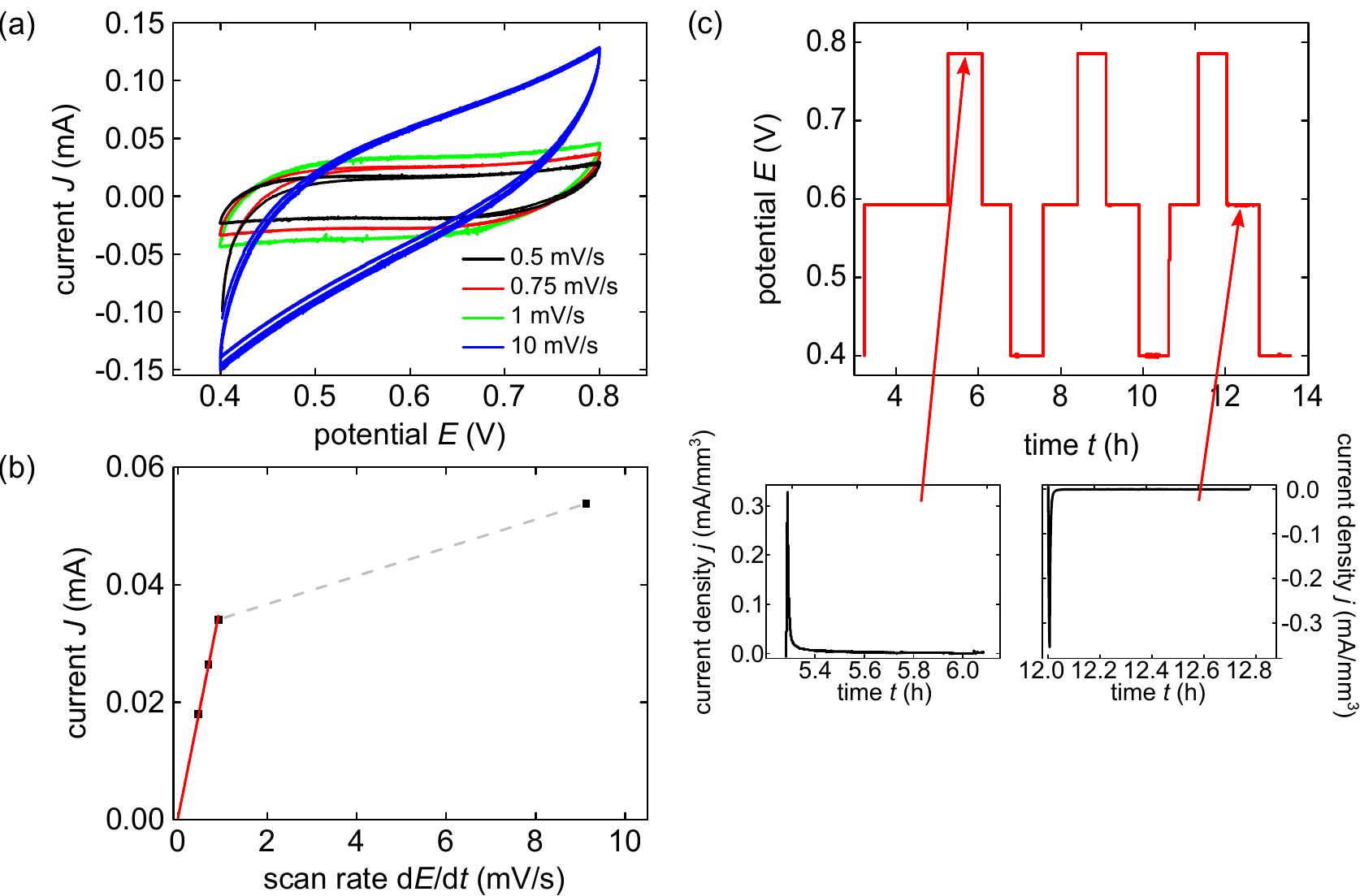}
		\caption{\textbf{Electrochemical characterisation of sample $\mathrm{S_{refl}}$.} (a) The graph depicts exemplary cyclic voltammetry measurements of the np-Si/PPy hybrid $\mathrm{S_{refl}}$ in $1\,\mathrm{mol\, l^{-1}}$ $\mathrm{HClO_4}$ electrolyte solution in the in situ electrochemical measurement cell. The current $J$ is plotted against the applied potential $E$ in the range of $0.4\,-\,0.8\,\mathrm{V}$, measured versus the standard hydrogen potential (SHE). The potential scan rate is increased from $0.5\,\mathrm{mVs^{-1}}$ to $10\,\mathrm{mVs^{-1}}$. (b) The graph depicts values for the np-Si volume specific current $j$, averaged in the potential range of $0.55\,-\,0.65\,\mathrm{V}$, plotted against increasing potential scan rates $\mathrm{d}E/\mathrm{d}t$ from $0.5\,\mathrm{mVs^{-1}}$ to $10\,\mathrm{mVs^{-1}}$. The red line indicates a linear regression of the first three data points which yields the capacitance $c_\mathrm{norm, refl}=(0.0293\pm0.0004)\,\mathrm{Fmm^{-3}}$ as the slope. (c) The upper graph depicts the course of the potential during the successive full $\theta-2\theta$ XRD measurements of the electrochemical actuation. The potential $E$ is stepped from $0.4\,\mathrm{V}$ over $0.6\,\mathrm{V}$ to $0.8\,\mathrm{V}$ and back and remains at each step for as long a XRD measurement takes. The smaller, lower Figures depict the course of the np-Si volume specific current $j$ during one potential step up (left) and down.}
		\label{Fig_Echem}
	\end{figure*}
	Figure \ref{Fig_Echem}(a) depicts a CV measurement in this potential range with scan rates of $0.5$, $0.75$, $1.0$ and $10.0\,\mathrm{mV/s}$. The course of the measurement conducted with $0.5\mathrm{mV/s}$ shows a quick conversion towards a constant current value of about $0.016\,\mathrm{mA}$ starting at the lower vertex point at $0.4\,\mathrm{V}$. After reversing the scan direction at the upper vertex point of $0.8\,\mathrm{V}$ the current decreases to a constant value of approximately $-0.019\,\mathrm{mA}$. Hence, defined peaks, characteristic for the presence of electrochemical reactions, in neither up nor down sweep are distinctively apparent. Furthermore, the charge transferred into the polymer on the up-sweep can be fully recovered on the down-sweep, as the associated currents are virtually identical bar their sign. Thus, the potential regime chosen here represents a reversible capacitive charging of the np-Si/PPy hybrid. The measurements conducted with scan rates of $0.75$, and $1.0\,\mathrm{mV/s}$ exhibit identical characteristics of the potential course with a proportionally increasing current level, albeit a small increase of the current towards the upper reversal point becomes visible. However, a scan rate of $10\,\mathrm{mV/s}$ leads to a noticeably difference. A constant current with the increasing potential is not achieved, instead the current is linearly increasing. Hence, a diffusion limitation of the anion charge carriers inside the polymer is present which prohibits an sufficiently fast movement to counterbalance the increasing potential for higher scan rates~\cite{Hamann2007}. The capacity of the sample can be obtained by the following procedure. The absolute current values between $0.55\,\mathrm{V}$ and $0.65\,\mathrm{V}$ are averaged for both scan directions and plotted versus their respective scan rate and normalized to the volume of the electrochemically active, np-Si layer. The plot is depicted in Figure \ref{Fig_Echem}(b). A linear regression, according to Equation \ref{Eq:capacity}, yields the capacitance $c$ according to Equation \ref{Eq:capacity}. Note, that the data point with a scan rate of $10\,\mathrm{mV/s}$ is omitted. The result of the sample measured in the reflection geometry reads $c_\mathrm{norm, refl}=0.0347/(A_\mathrm{refl}\cdot t_\mathrm{np-Si})=(0.0293\pm0.0004)\,\mathrm{Fmm^{-3}}$, a value in good agreement with literature~\cite{Brinker2020}. All in all, the here ascertained electrochemical behavior is typical for a hybrid material of PPy and np-Si and has been reported before. Thus, the investigated material exhibits a desired electrochemical behavior and the anion incorporation can be controlled via the applied potential $E$. Therefore, the lattice mismatch under the influence of the potential controlled electrochemical actuation can be investigated.\\
	For $\mathrm{S_{trans}}$, a CV measurement is conducted with the same parameters as the CV for the reflection geometry. The result is depicted in the supplemetary materials in Figure S5(a)\cite{Supp2022}. It also shows the same capacitive features. The analysis of the capacity, cf. Figure S5(b), results in a value of $c_\mathrm{norm, trans} =(0.0308\pm0.0006)\,\mathrm{Fmm^{-3}}$, which is in good agreement with the sample measured in reflection geometry.\\
	Another type of measurement is a constant potential with the simultaneous measurement of $j$ and $Q$. The potential range is divided into discrete potential steps of $0.4$, $0.6$ and $0.8\,\mathrm{V}$. These potential steps can be approached one after another and the CV is mimicked in a step-like fashion. At each potential step a full $\theta$-$2\theta$ measurement can be measured. The applied potential and the response of the current at one upward- and one downward potential step is depicted in Figure \ref{Fig_Echem}(c). The current changes instantly on a change of the potential. If the potential is stepped from a lower to a higher value, the current instantly increases and then asymptotically approaches $0$. The current course therefore signifies an incorporation of charge carriers into the sample. The reaction of the current for a step of a higher to a lower potential is of opposite nature. Integrating the current for these potential steps yields the exchanged charge $Q$.\\
	For an assessment of the kinetics of the electrochemical actuation process, a square potential, with instant changes from a lower- to a higher potential step and vice versa, is applied.\\
	\subsection{Transmission electron microscopy of porous silicon}
\begin{figure*}
		\centering
		\includegraphics[width=\textwidth]{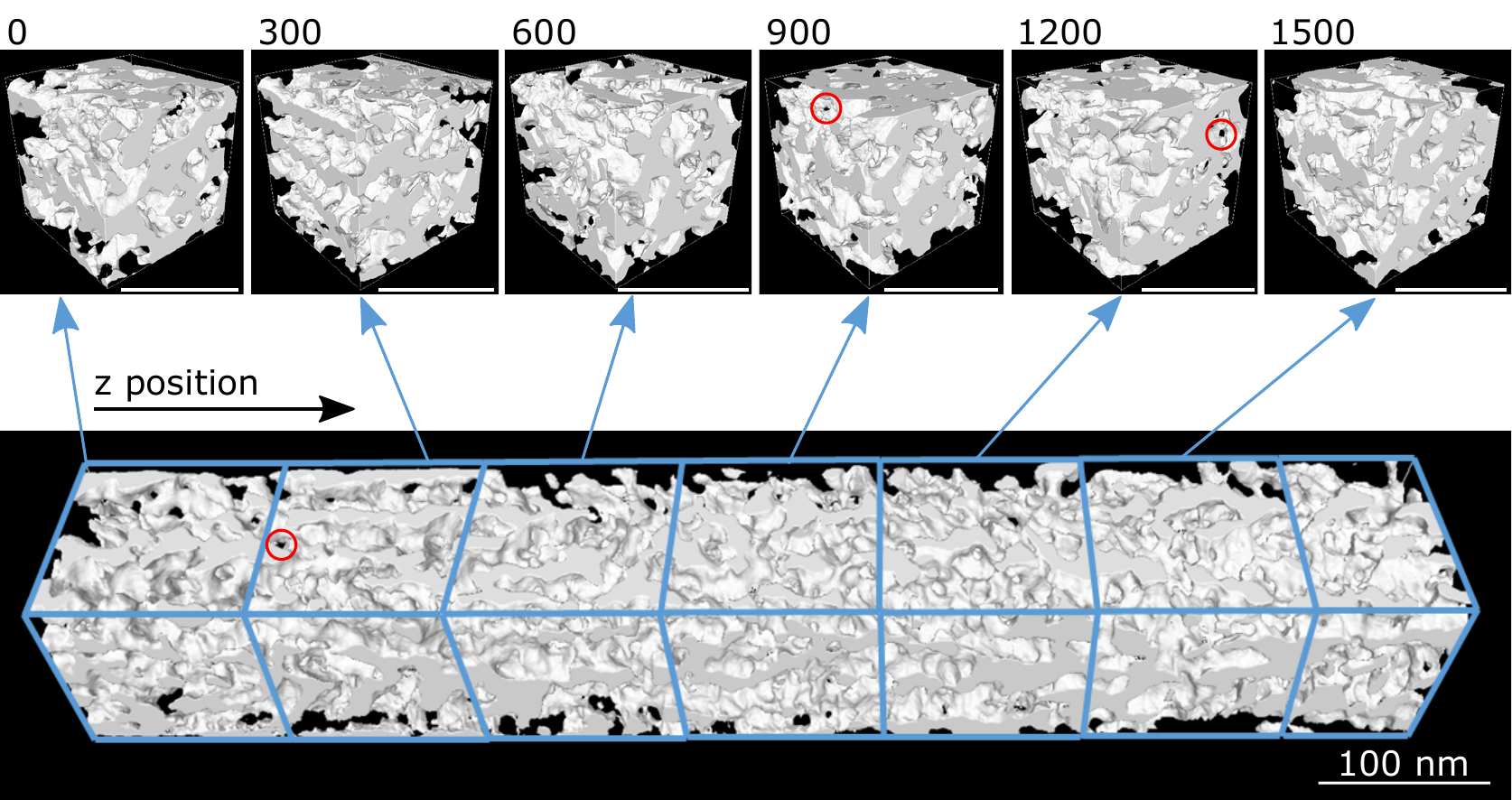}
		\caption{\textbf{Micromechanical formation of the simulation model.} Result of 3D TEM tomography after segmentation. The position in z-direction corresponds to the position in the image-stack, where z moves from top to bottom. The scale bar in each picture has a length of $100\,\mathrm{nm}$. Exemplary interconnections between main pores indicated by red circles.}
		\label{Fig_TEMstack}
	\end{figure*}

	A needle-shaped tomography sample is prepared with an focused-ion beam (FIB, FEI Helios G3 UC) machine using a $30\,\mathrm{keV}$ $\mathrm{Ga}^{+}$ ion beam and transferred on a lift-out needle stub via in-situ lift-out technique. The diameter of the tomography sample is approximately $100\,\mathrm{nm}$.
	A TEM (FEI Talos F200X) equipped with a high brightness Schottky-FEG (X-FEG) acquires a tilt series of the needle. The microscope parameters are $200\,\mathrm{kV}$ acceleration voltage, $50\,\mathrm{pA}$ beam current and a camera length of $98\,\mathrm{mm}$ for the scanning transmission electron microscopy (STEM) high angle annular dark-field (HAADF) detector. The resolution of the images is $2048 \times 2048$ pixels with a pixel size of $384\,\mathrm{pm}$. The tilt range was +80 to -80 degrees with an increment of 2 degrees resulting in 81 images. The image registration and volume reconstruction was performed with the software Inspect3D (Thermo Fischer Scientific USA) using the EM algorithm~\cite{Lange1984}. The reconstructed volume is binarized using the intensity histogram of a representative sample subvolume using Otsu’s method~\cite{Otsu1979} implemented in the software Fiji.\\
	Figure \ref{Fig_TEMstack} shows that np-Si contains a complex three-dimensional network of pores, which appear highly random and indicates interconnections between the pores. The porosity is slightly decreasing from top to bottom by about 4\%. The segmentation is calibrated for the cubic slice of $300 \times 300 \times 300$ voxel at position $z=900$ to a relative density of $\Phi=45.7\,\%$. Moreover, it can be seen in the single segments, that the main pores exhibit interconnections, which would confirm recent results in literature~\cite{Shchur2022,Bossert2021,Kondrashova2017}. The main pores in the tomography are in a range of $7-25\,\mathrm{nm}$, but are not further evaluated here. 
	\subsection{Micromechanical simulation}
\begin{figure*}
		\centering
		\includegraphics[width=\textwidth]{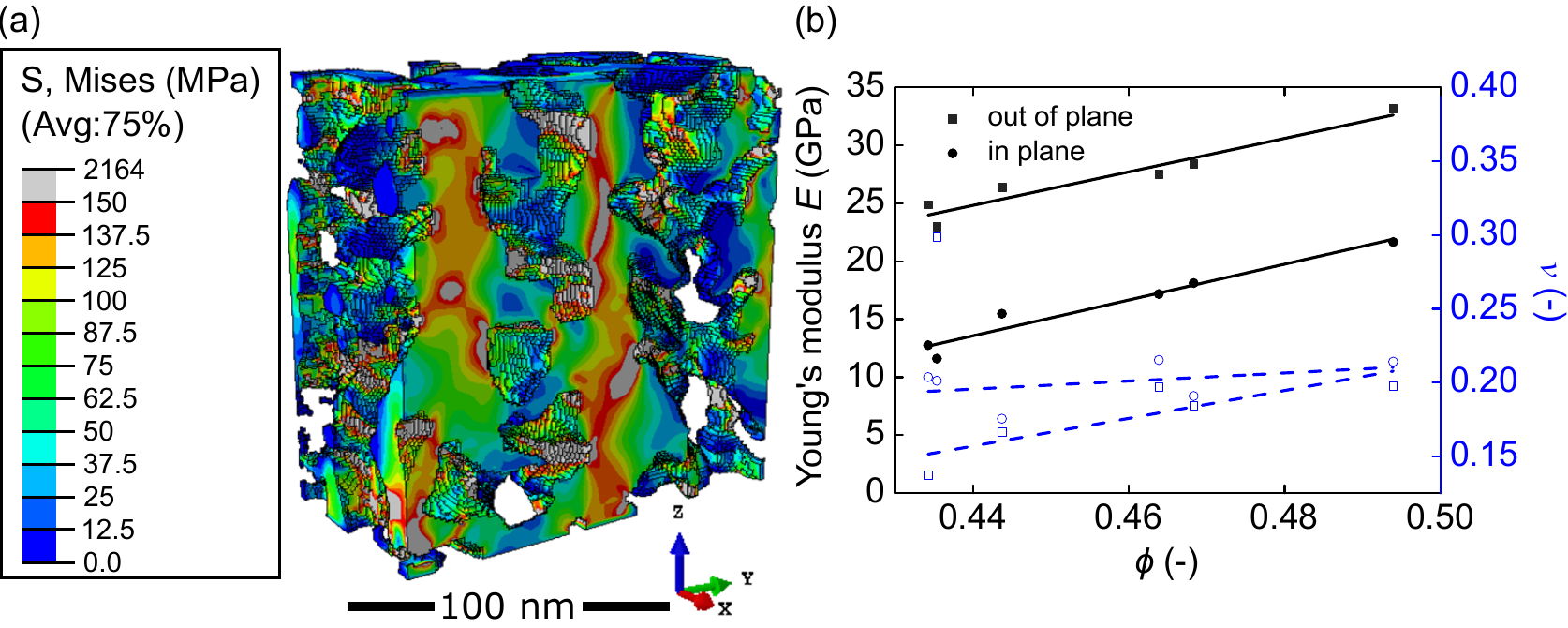}
		\caption{\textbf{Fabrication of the Simulation model.} (a) Microstructure of the RVE at position $z=900$ with mises stress under external load in $z$-direction displayed for cut through the mid-plane of the FE-voxel model. The scale bar represents $100\,\mathrm{nm}$. (b) Predicted mechanical properties as function of solid fraction $\phi$.}
		\label{Fig_RVE}
	\end{figure*}

	For the micromechanical simulation, the segmented volume at $z=900$ shown in Fig. \ref{Fig_TEMstack} is translated into a cubic voxel model. First, the voxel structure is coarsened by a factor five without changing the total solid fraction. This leads to a representative volume element (RVE) with a voxel resolution of $150 \times 150 \times 150$ that consists of 3.375 million elements when the RVE is completely filled. The silicon phase in the RVE is modeled with orthotropic elasticity according to a standard $(100)$ silicon wafer, which is $[110]$, $[110]$, $[001]$ with $E_1=E_2=169\,\mathrm{GPa}$, $E_3=130\,\mathrm{GPa}$, $\nu_{23}=0.36$, $\nu_{31}=0.28$, $\nu_{12}=0.064$, and $G_{23}=G_{32}=79.6\,\mathrm{GPa}$, $G_{12}=50.9\,\mathrm{GPa}$~\cite{Hopcroft2010}. 
	The remaining pore space can be either be left empty or filled with PPy, modeled with isotropic elastic with $E_\mathrm{p}=500\,\mathrm{MPa}$ and $\nu_\mathrm{p}=0.35$~\cite{Brinker2020}. It should be noted, that these values are valid for a PPy film and the polymer within the pores might differ, due to the confinement.\\
	With the np-Si model, shown in Fig. \ref{Fig_RVE}(a), the macroscopic mechanical properties are determined in a first step by a small tensile strain of $0.1\%$ that is subsequently applied in all three coordinate directions. The von Mises stress distribution in Fig. \ref{Fig_RVE}(a) is visualized for the mid cross-section for loading in $z$-direction (out-of-plane). Due to the random pores that are penetrating the silicon 'walls', the load transmission in $z$-direction is very inhomogeneous and the assumption of a $2\,\sfrac{1}{2}\,$D model as it was employed in~\cite{Brinker2020} does not hold. Consequently, the out-of-plane direction ($z$-direction) turns out to be much more compliant. For a porosity of $\approx 50\,\%$ that is perfectly aligned in $z$-direction, the macroscopic Young's modulus should be $\approx 65\,\mathrm{GPa}$, whereas the simulation results shown in \ref{Fig_RVE}(c) predict about half of this value, ranging from $23\,\mathrm{GPa}$ to $33\,\mathrm{GPa}$ with increasing depth position $z$ and solid fraction $\phi$. As shown in \ref{Fig_RVE}(c), the macroscopic stiffness changes linearly with the solid fraction $\phi$. The in-plane stiffness is shifted by $\approx-10\,\mathrm{GPa}$, relative to the out-of-plane stiffness, and reaches at the lower end the value of $\approx10\,\mathrm{GPa}$ that has also been measured in macroscopic experiments~\cite{Brinker2020}. 
	Interestingly, the Poisson's ratio measured for the out-of-plane direction follows the same linear trend, while the value for loading in the two in-plane directions remains fairly constant $\nu_{x,y}=0.2$. In the following, we use the RVE derived from the segmented TEM tomography at the mid position at $z=900$ for the calibration of further model parameters.\\
	%
	%
	%
	For incorporating surface effects that are detected in the XRD measurements, the interfaces between silicon and PPy are covered with shell elements~\cite{Huber2021}. They serve to introduce a surface strain that emulates the expansion of the free surface in the as-prepared and liquid infiltrated state. To this end, all faces of voxels shared by two different phases, additionally share a four node shell element of a thickness of $10^{-3}$ of the RVE size. The elastic properties of the shell elements are chosen with $E=169\,\mathrm{GPa}$ and $\nu=0.36$, such that only small expansions are sufficient to insert a detectable strain into the surface of the silicon network. The coefficient of expansion of the interface elements can be adjusted to simulate the surface strain in the silicon in the as-prepared and liquid infiltrated states. Here, the as-prepared state is arbitrarily represented by a temperature increase of $1\,\mathrm{K}$ in the model.\\
	%
	%
	Due to the thin and compliant porous layer on top of the bulk silicon wafer, in-plane macroscopic strains can be neglected. This leads to boundary conditions with zero normal displacements at the bottom and the four sides of the RVE shown in Fig. \ref{Fig_RVE}. For later visualization of the swelling of the PPy phase, the top face is allowed to move, but it is forced to remain plane.
	\section*{Results}
	\subsection*{Out-of-Plane Strain}
	\begin{figure*}
		\centering
		\includegraphics{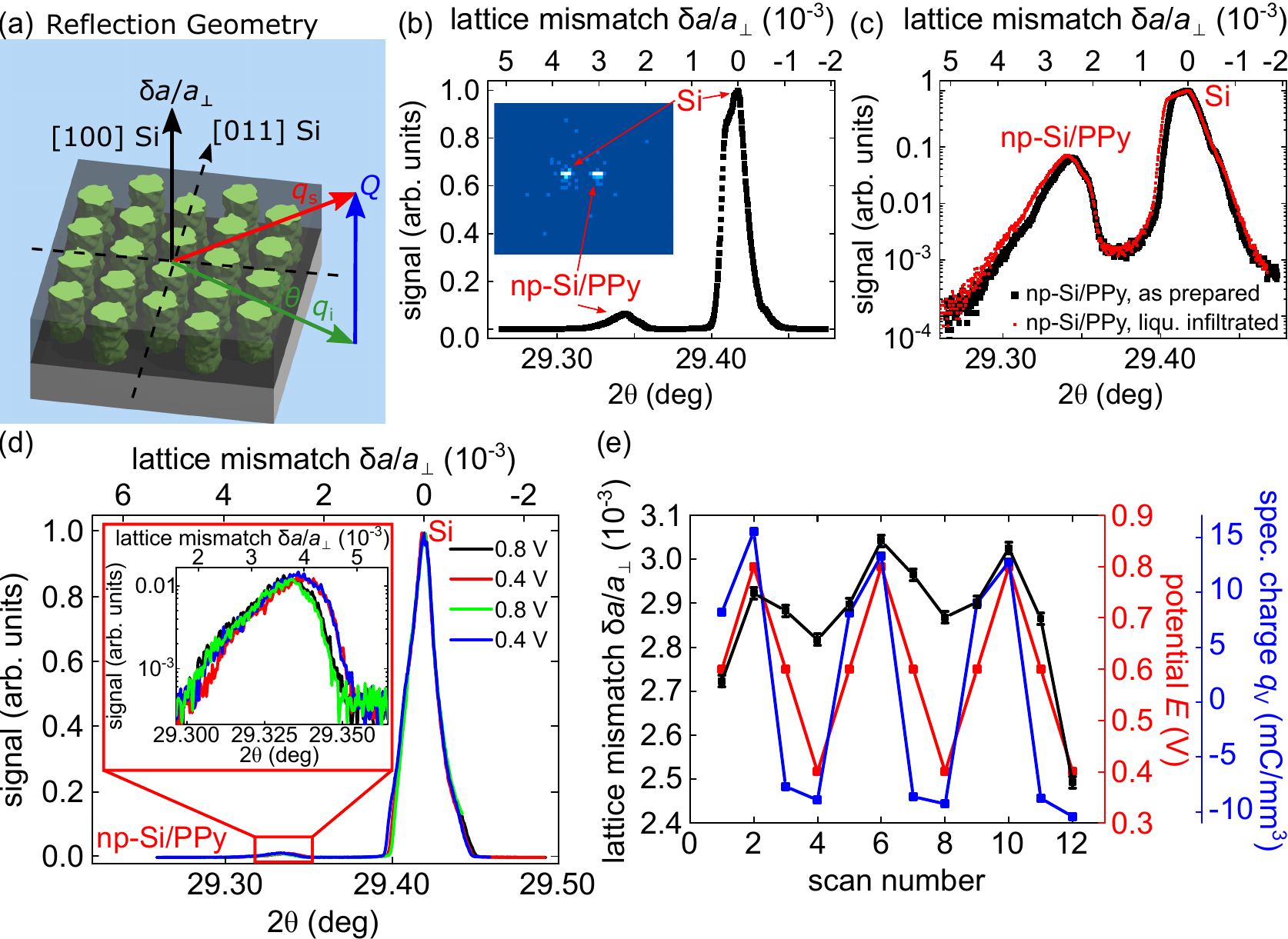}
		\caption{\textbf{In operando X-ray diffraction measurement in reflection geometry.} (a) Diffraction geometry of the np-Si (dark grey) PPy (green) hybrid attached to bulk silicon (light grey), with incident beam $q_\mathrm{i}$, diffracted beam $q_\mathrm{s}$, the transferred wave vector $Q$, crystallographic directions and the axis of the probed lattice mismatch $\delta a/a_\perp$. (b) Radial $\theta-2\theta$ scan of the (400) Bragg peak in the out-of-plane sample direction in the as-prepared state without an electrolyte solution. The ordinate is in linear scale, the abscissa shows angle $2\theta$ at the bottom and lattice mismatch $\delta a/a_\perp$ with respect to the main silicon peak at the top, respectively. The inset depicts the actual detector picture of the Si and np-Si/PPy signal. (c) Depicted is again the (400) $\theta-2\theta$ scan of the as-prepared sample (black square symbols). The ordinate is in logarithmic scale in contrast to (b). An equal Radial $\theta-2\theta$ scan of the (400) Bragg peak in the out-of-plane direction is shown on the np-Si/PPy hybrid infiltrated with $1\,\mathrm{mol\,l^{-1}}$ perchloric acid (red square symbols). The graphic yields a direct comparison between the two different sample states -- as-prepared versus electrolyte-infiltrated. (d) Potential dependent X-ray diffraction measurements. The figure depicts $\theta-2\theta$ scans of the (400) Bragg peak in the out-of-plane direction of the infiltrated np-Si/PPy hybrid infiltrated by $1\,\mathrm{mol\,l^{-1}}$ perchloric acid for applied potentials of $E=0.4\,\mathrm{V}$ and $E=0.8\,\mathrm{V}$, respectively. The inset represents a zoom in the Bragg peak that is caused by the np-Si layer. (e) The resulting lattice mismatch $\delta a/a_\perp$ of the (400) Bragg peak in the out-of-plane direction and the np-Si volume specific charge $q_\mathrm{V}$ in dependence of the applied potential $E$ at each respective step.}
		\label{Fig_XRD-out-ofplane}
	\end{figure*}
	The first X-ray diffraction experiment is performed in reflection geometry with an empty cell, i.e., no electrolyte solution is filled. Hence, the lattice mismatch of the as synthesized sample is characterized. The respective measurement is presented in Figure \ref{Fig_XRD-out-ofplane}(b). The ordinate is displayed linearly and normalized by the intensity of the peak at $2\theta\approx29.42\,\mathrm{^{\circ}}$. The presence of two distinct peaks is a characteristic X-ray diffraction feature of a crystalline np-Si layer attached to a bulk silicon substrate. The main peak corresponds to the bulk silicon's (400) lattice planes. At a smaller angle of $2\theta\approx29.34\,\mathrm{^{\circ}}$ another distinct peak is clearly visible albeit with one order of magnitude lower intensity. It is characteristic of the (400) planes in the np-Si layer. The difference in intensity of approximately $90\%$ can be ascribed to the thickness variance of the layers. Whereas the bulk silicon layer has a thickness of $500\,\mathrm{\upmu m}$, the thickness of the np-Si layer only amounts to $t=25.4\,\mathrm{\upmu m}$ while only having roughly half the density of bulk silicon due to its porosity. Since the angle $2\theta$ of the np-Si layer peak is at a smaller value than bulk silicon's, the lattice mismatch $\delta a/a_\perp$ is positive. Thus, the lattice of the np-Si layer is expanded in the $[100]$ direction. To obtain a quantitative result for the lattice mismatch, both $2\theta$ peaks need to be approximated by a suitable fit function. For this purpose the fit program LIPRAS is used~\cite{LIPRAS2017}. As can be seen in Figure \ref{Fig_XRD-out-ofplane}(b) the bulk silicon $2\theta$ peak has no symmetric shape resembling a Lorentz or Gauss profile. Rather, the main peak is distorted by a broad shoulder at a slightly smaller angle and a foot at higher angles than the main peak. This distortion is further debated in the discussion. To fit the course of the $2\theta$ signal, four pseudo-Voigt functions are utilized. Thereby, the main peak is fitted with one while the broad shoulder at smaller angles is covered by one independent fit function, as is the foot of the signal at slightly higher $2\theta$ values with two fits. All here presented $\theta$-$2\theta$ measurements are evaluated by this fitting protocol. The fit in total approximates the silicon peak well, as can be seen in Figure S6 in the supplementary materials\cite{Supp2022}, and has an adjusted $r^2$-value of $99.36\%$. For the further evaluation of the lattice mismatch, only the peak position of the corresponding main peak fit will be considered. The result for the peak presented in Figure \ref{Fig_XRD-out-ofplane}(b) is $2\theta=(29.4159\pm0.0003)\,^{\circ}$. The angular uncertainty is discussed in the methods section. The main silicon peak exhibits a small deviation to the value of $2\theta=29.3934\,^{\circ}$, according to Equation \ref{Eq_Bragg} in the methods section. The deviation can be explained by a possible misalignment of the sample with regard to the experimental cell and the incident beam. However, as already emphasized, the measurement method to determine $\delta a/a_\perp$ only requires the relative difference of the bulk- and the porous silicon diffraction peaks. Hence, the np-Si $2\theta$ peak is evaluated with an equal fitting procedure which results in a value of $2\theta=(29.3437\pm0.0003)\,^{\circ}$. Thus, the two $2\theta$-peaks of porous- and bulk silicon result in a lattice mismatch of $\delta a/a_\perp=+(2.40\pm0.01)\cdot10^{-3}$. The equation for the lattice mismatch is given in the methods section as Equation \ref{Eq_da/a}.\\
	In a next step perchloric acid is added and the counter (CE) and reference electrodes (RE) are inserted into the cell and connected to the potentiostat, to perform electrochemical measurements. After the filling of the cell with perchloric acid, a $\theta$-$2\theta$ scan is performed to assess whether a change in the lattice constant of the np-Si layer occurs through an expansion driven by the absorption of the electrolyte solution. The respective measurement can be found in Figure \ref{Fig_XRD-out-ofplane}(c). The measurement reveals a lattice mismatch of $\delta a/a_\perp=+(2.48\pm0.01)\cdot10^{-3}$ for the sample infiltrated with the electrolyte solution. Thus, the difference in lattice mismatch between the as-prepared sample and the solution infiltrated sample is small but detectable and amounts to $\Delta (\delta a/a_\perp)=+(8\pm1)\cdot10^{-5}$. That means, the lattice expands even further in the out-of-plane direction upon the intake of the electrolyte solution.
	\subsection*{Out-of-Plane Strain - Electrochemical Strain Dependence and Kinetics}
	A $\theta$-$2\theta$ scan is conducted for each respective potential step. Two exemplary $\theta-2\theta$ measurements with applied potentials of $0.4\,\mathrm{V}$ and $0.8\,\mathrm{V}$ are depicted in Figure \ref{Fig_XRD-out-ofplane} (d).
	\begin{figure*}
		\centering
		\includegraphics{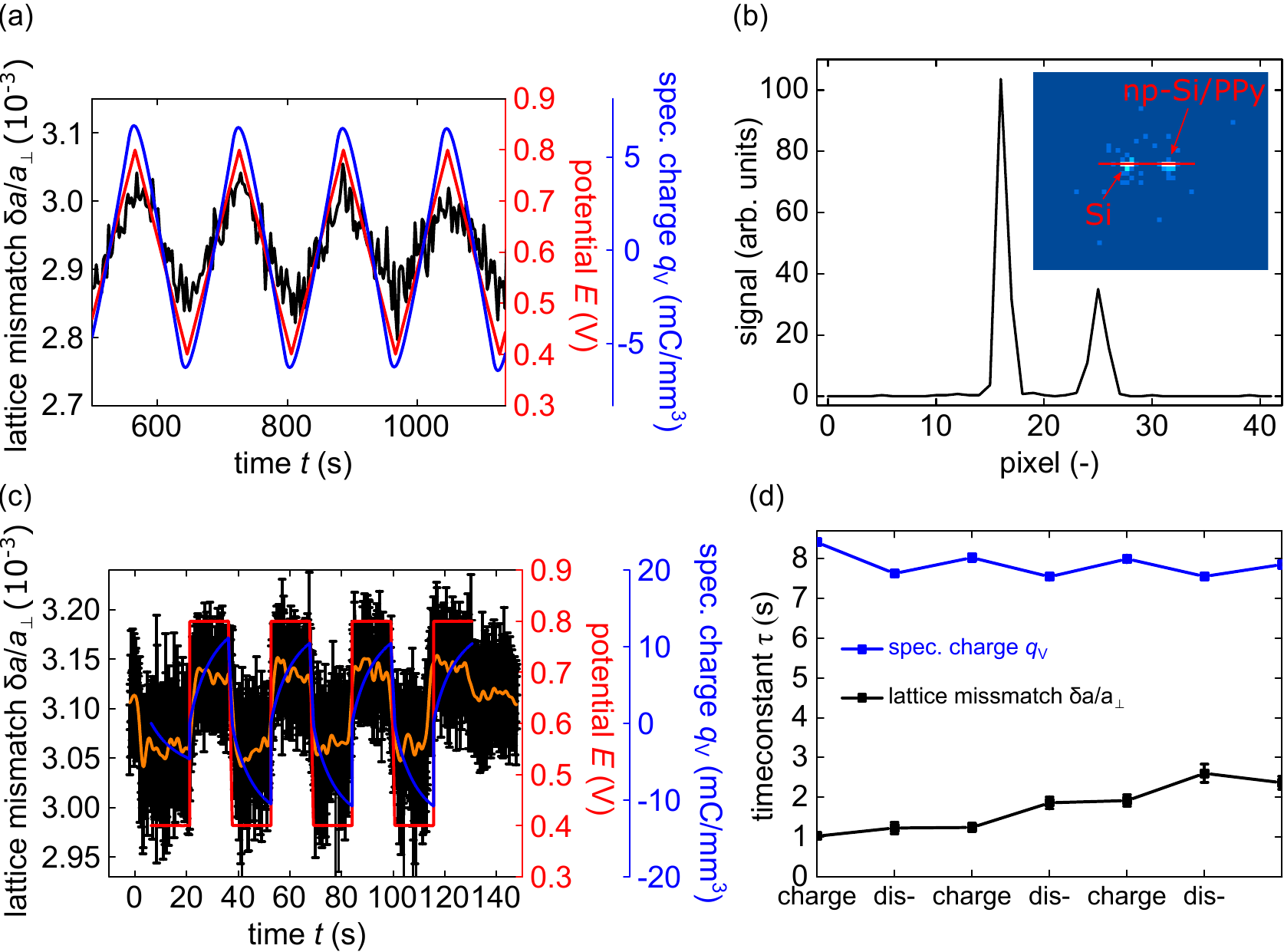}
		\caption{\textbf{Actuation kinetics of the lattice mismatch.} (a) Lattice mismatch $\delta a/a_\perp$ of the (400) Bragg peak in the out-of-plane direction and the volume specific charge $q_\mathrm{V}$ of np-Si in dependence of the applied, linearly changing CV potential $E$. (b) Exemplary profile of the (400) Bragg peaks in the out-of-plane direction of the np-Si and the bulk silicon layer respectively. The inset gives the actual detector picture from where the profile is extracted with intensity normalized to the Si main peak. (c) A step-coulombmetry measurement is performed in $1\,\mathrm{mol\, l^{-1}}$ $\mathrm{HClO_4}$ electrolyte solution for the determination of the electrochemical actuation kinetics of the np-Si/PPy hybrid. The applied potential $E$ (red) is changed in an instant fashion from $0.4\,\mathrm{V}$ to $0.8\,\mathrm{V}$ and vice versa. The incorporated volume specific charge $q_\mathrm{V}$ (blue) and the resulting lattice mismatch $\delta a/a_\perp$ (black, with the average in orange) are depicted versus time $t$ along with the potential $E$. (d) The time constants $\tau$ of the $\delta a/a_\perp$ and $q_\mathrm{V}$ reaction to the square potential depicted in (c).}
		\label{Fig_time}
	\end{figure*}
	The main silicon peak does barely change its center peak position with applied potential. The np-Si/PPy peak on the other hand experiences a clearly visible shift in dependence of the applied potential, as can be seen in the inset of Figure \ref{Fig_XRD-out-ofplane}(d). The lattice mismatch $\delta a/a_\perp$ can be calculated from the two peaks for all potential steps. The result is shown in Figure \ref{Fig_XRD-out-ofplane}(e) along with the applied potential $E$ and the consumed charge per volume of the porous silicon layer $q_\mathrm{V}$. The lattice mismatch follows the potential and the accumulated charge in phase. Increasing the potential results in an incorporation of anions, which leads to an increase in lattice mismatch and, vice versa, decreasing the potential leads to an emission of charge carriers and thereby a smaller lattice mismatch. Hence, to highlight this, an increasing potential signifies an expansion of the np-Si lattice in the out-of-plane direction and a potential reduction indicates a contraction of the lattice. The lattice mismatch $\delta a/a_\perp$ varies between approximately $+0.0028$ and $+0.0030$. The amplitude of the lattice mismatch actuation averaged over the two center cycles (scan 2 to 10) amounts to $A_{\delta a/a_\perp}=(1.50\pm0.04)\cdot10^{-4}$. All in all, the observed reaction of the lattice upon an applied potential is remarkably reversible at the lower- and upper vertex point and in total very reproducible.\\
	The reproducibility and the linearity of the lattice mismatch dependence on potential is investigated in a further measurement, where the change in intensity at one angle is related to the lattice mismatch. Therefore, the angle $2\theta$, and the incident angle $\theta$ accordingly, is fixed to the right flank of the np-Si peak at half the intensity of the peak. Thus, a shifting of the peak has the highest effect of the change in intensity at this edge position. The angle is set to $2\theta=29.3443^{\circ}$ and is left constant for the measurement, during which only the intensity is summed in intervals of $1\,\mathrm{s}$. Hence, only a shift in intensity is measured. To compare it, the shift in intensity is converted into a shift in lattice mismatch $\delta a/a_\perp$ by the following procedure. The fit parameters of the latest preceding full $\theta$-$2\theta$ measurement's fit are utilized. Then, the $\theta$ angle is adjusted, while the other determined fit parameters are kept constant, until the measured intensity at the edge position can be reproduced. Hence, the peak of the np-Si layer's Bragg peak position can be inferred. As the peak shape stays the same at different potentials, see inset Figure \ref{Fig_XRD-out-ofplane}(d), this procedure is viable.\\
	Simultaneously to the recording of the intensity, a CV measurement is conducted. The potential range is $0.4-0.8\,\mathrm{V}$. It is the same as in the respective step-wise method and thus renders these measurements comparable. The course of the lattice mismatch in response to the linearly changing CV potential is depicted in Figure \ref{Fig_time}(a) along with the volume specific charge $q_\mathrm{V}$. The lattice mismatch follows both $E$ and $q_\mathrm{V}$ in a linear and reversible fashion. The lattice mismatch has an amplitude of $A_{\delta a/a_\perp}=(1.55\pm0.03)\cdot10^{-4}$ and is in excellent agreement with the amplitude determined by the step-wise method.\\
	The kinetics of the actuation process are investigated in an analogous manner. A square potential from $0.4-0.8\,\mathrm{V}$ instead of a linearly changing CV potential is applied and the response of the lattice mismatch $\delta a/a_\perp$ and the volume specific charge $q_\mathrm{V}$ is probed. The angle $2\theta=29.3342^{\circ}$ is therefore fixed at the np-Si peak position, to maximize the signal. The temporal resolution has to be maximized, due to the fast lattice response upon an instant potential change. The former measurement technique does not provide the necessary temporal resolution, i.e. summing over the central detector region, see methods section. Thus, the lattice mismatch is determined by evaluating the actual detector pictures, recorded with a resolution of $0.1\,\mathrm{s}$. An exemplary detector picture is shown in Figure \ref{Fig_time}(b) in the inset. The bulk silicon peak is visible on the left of the detector picture and to the right the peak is caused by the np-Si layer. A profile of the peaks is fitted, to determine the pixel of the peak positions. These pixel positions can then be assigned $2\theta$ values by their detector pixel number relative to the detector center and the distance from the sample. Finally, the respective $2\theta$ angle can be transformed into a value for the lattice mismatch and the respective $2\theta$ value of the detector center. By this means, the temporal resolution of the measurement is enhanced and one lattice mismatch value for each $0.1\,\mathrm{s}$ time step can be determined. Figure \ref{Fig_time}(c) depicts the resulting measurement along applied potential and accumulated charge. The lattice mismatch reacts to the abrupt potential changes in a similar square like fashion. The accumulated charge $q_\mathrm{V}$ on the other hand shows a distinctly different behavior as it takes significantly longer to settle to a plateau. Exponential functions are used to determine the time constants $\tau$ of both lattice mismatch and volume specific charge. $\tau$ yields the speed of the respective processes, i.e., the charging and discharging of the PPy and the accompanying reaction of $\delta a/a_\perp$. The results are depicted in Figure \ref{Fig_time}(d). The reaction of the lattice, with values between $1$ and $3\,\mathrm{s}$, is faster than the charge movement, which exhibits values for $\tau$ of about $7$ to $8\,\mathrm{s}$. It can be seen, that a slight drift is present in the charging times.
	\subsection*{In-Plane Strain}
\begin{figure*}
		\centering
		\includegraphics{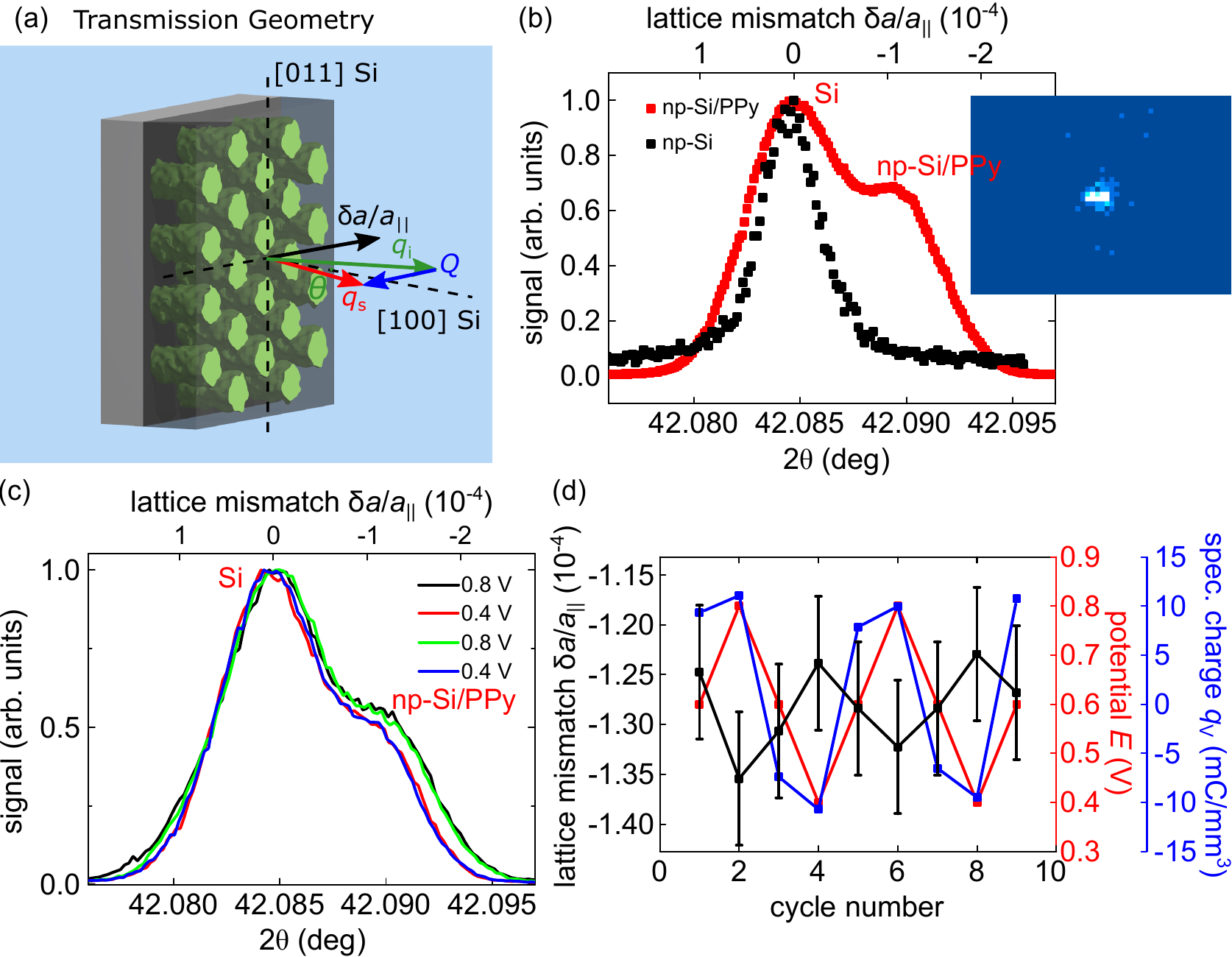}
		\caption{\textbf{In operando X-ray diffraction measurements in transmission geometry.} (a) Diffraction geometry of the np-Si (dark grey) / PPy (green) material attached to bulk silicon (light grey), with incident beam $q_\mathrm{i}$, diffracted beam $q_\mathrm{s}$, the wave vector transfer $Q$, crystallographic directions and the axis of probed lattice mismatch $\delta a/a_{||}$. For the in-plane mismatch in the transmission geometry, the incident beam hits the backside of the sample, is transmitted through the sample and leaves the sample under an angle $\theta$. (b) $\theta-2\theta$ XRD measurements of the (440) Bragg peak in the in-plane direction of the np-Si/PPy hybrid sample infiltrated by $1\,\mathrm{mol\,l^{-1}}$ perchloric acid in comparison with a plain, unfilled np-Si sample. The abscissa shows angle $2\theta$ at the bottom and lattice mismatch $\delta a/a_{||}$ with respect to the main silicon peak at the top, respectively. The inset depicts the actual detector picture of merged Si and np-Si/PPy signal. (c) $\theta-2\theta$ XRD measurements of the (440) Bragg peak in the in-plane direction of the electrolyte infiltrated np-Si/PPy hybrid sample in dependence of the applied potentials $E=0.4\,\mathrm{V}$ and $E=0.8\,\mathrm{V}$. (d) The resulting lattice mismatch $\delta a/a_{||}$ of the (440) Bragg peak in the in-plane direction and the np-Si volume specific charge $q_\mathrm{V}$ in dependence of the applied potential $E$ at each respective step.}
		\label{Fig_XRD_in-plane}
	\end{figure*}

\begin{figure*}
		\centering
		\includegraphics{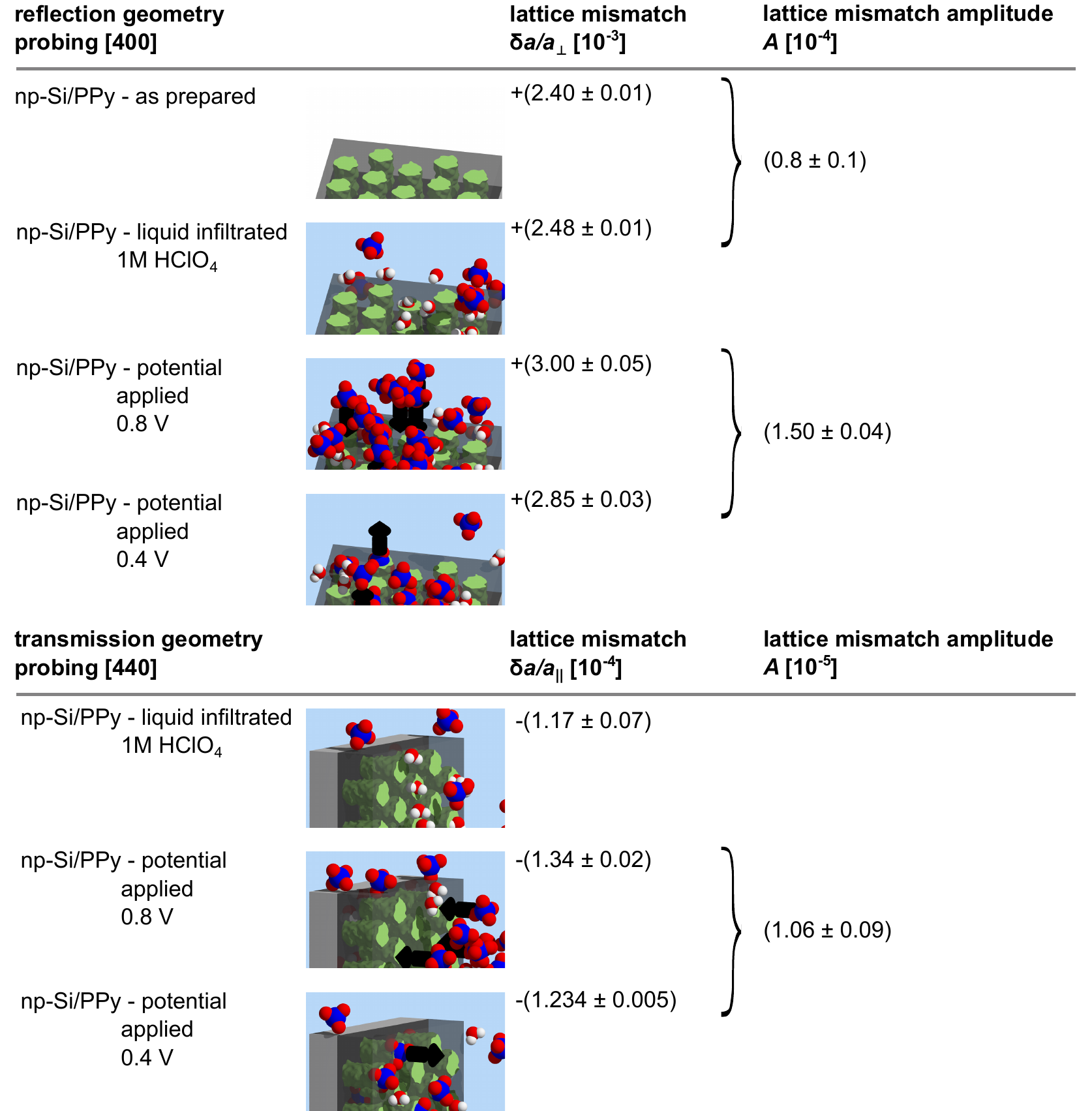}
		\caption{\textbf{Overview of overall results.} Overview of the results of the measurements conducted in reflection and transmission geometry with the different states of the np-Si/PPy hybrid - from as-prepared over liquid-infiltrated to two different applied potentials $E=0.4\,\mathrm{V}$ and $E=0.8\,\mathrm{V}$.}
		\label{Fig_Table}
	\end{figure*}

	Figure \ref{Fig_XRD_in-plane}(b) depicts a $\theta-2\theta$ measurement in transmission geometry, or rather a $(\theta-90^{\circ})-2\theta$ measurement, from approximately $2\theta\approx42.0800^{\circ}$ to $42.0950^{\circ}$ for both, a sample with a np-Si layer filled by PPy and an unfilled np-Si layer. 
	The respective measurements are conducted in the cell already filled with 1 molar $\mathrm{HClO_4}$ electrolyte solution. The np-Si layer without the polymer filling only shows one clear peak at $2\theta=42.0844^{\circ}$. By contrast, the np-Si layer with the polymer filling exhibits an additional small second peak in the right edge of the peak at $2\theta=42.0899^{\circ}$. This second peak can be ascribed to the effect of the polymer filling, which apparently causes significant strains on the lattice alone. As the peak of np-Si is at a slightly larger angle than the bulk silicon peak, the resulting lattice mismatch is negative. That means, the lattice is contracted in the in-plane direction. The two peaks of the $\theta-2\theta$ measurement are fitted by two pseudo-Voigt functions in the same manner as the peaks in reflection geometry. The resulting lattice mismatch amounts to $\delta a/a_{||}=-(1.17 \pm 0.07)\cdot10^{-4}$ and is thereby an order of magnitude smaller in absolute numbers than the lattice mismatch in the out-of-plane direction.\\
	To probe the lattice mismatch in the in-plane direction, it is ascertained that the sample behaves as desired electrochemically. In the subsequent measurement the potential $E$ is also controlled with steps at $0.4\,\mathrm{V}$, $0.6\,\mathrm{V}$ and $0.8\,\mathrm{V}$. Exemplary $\theta-2\theta$ measurements at the lower- and upper vertex point $0.4\,\mathrm{V}$ and $0.8\,\mathrm{V}$ are shown in Figure \ref{Fig_XRD_in-plane}(c). A clear shift to smaller values of $2\theta$, and thus a larger lattice mismatch, of the np-Si peak with respect to the bulk silicon peak for the higher voltage of $0.8\,\mathrm{V}$ is observable.\\
	Additionally, at each of the potential steps, the accumulated charge is acquired along with the lattice displacement resulting from the full $\theta-2\theta$ measurements. The respective data is plotted in Figure \ref{Fig_XRD_in-plane}(d). The course of the volumetric charge replicates the measurement conducted in the reflection geometry, depicted in Figure \ref{Fig_XRD-out-ofplane}(e), well. $q_\mathrm{V}$ oscillates in both measurements between approximately $-0.010\,\mathrm{mm^3C^{-1}}$ and $+0.012\,\mathrm{mm^3C^{-1}}$. The course of the lattice mismatch $\delta a/a_{||}$ in-plane, on the other hand, has an opposite dependency on $E$ and $q_\mathrm{V}$ compared to the out-of-plane direction. Here, an increase of the potential leads to an even larger decrease of $\delta a/a_{||}$ to more negative values, meaning that the lattice contracts further. Vice versa, a decrease in potential expands the lattice as $\delta a/a_{||}$ increases. Moreover, the error bars are larger compared to the measurement in reflection geometry. The amplitude of the lattice mismatch actuation in plane is $A_{\delta a/a_{||}}=(1.06\pm0.09)\cdot10^{-5}$ and is thus an order of magnitude smaller than for the out-of-plane direction.\\ 
	In summary, the the lattice mismatch in the in-plane- as well as the out-of-plane direction can be successfully determined, as well as the shifts due to applied potentials. The results are summarized in Figure \ref{Fig_Table}.
	Furthermore, the kinetics of the lattice mismatch in the out-of-plane direction are measured.

	\subsection*{Micromechanical Simulation of Actuation}
\begin{figure*}
		\centering
		\includegraphics{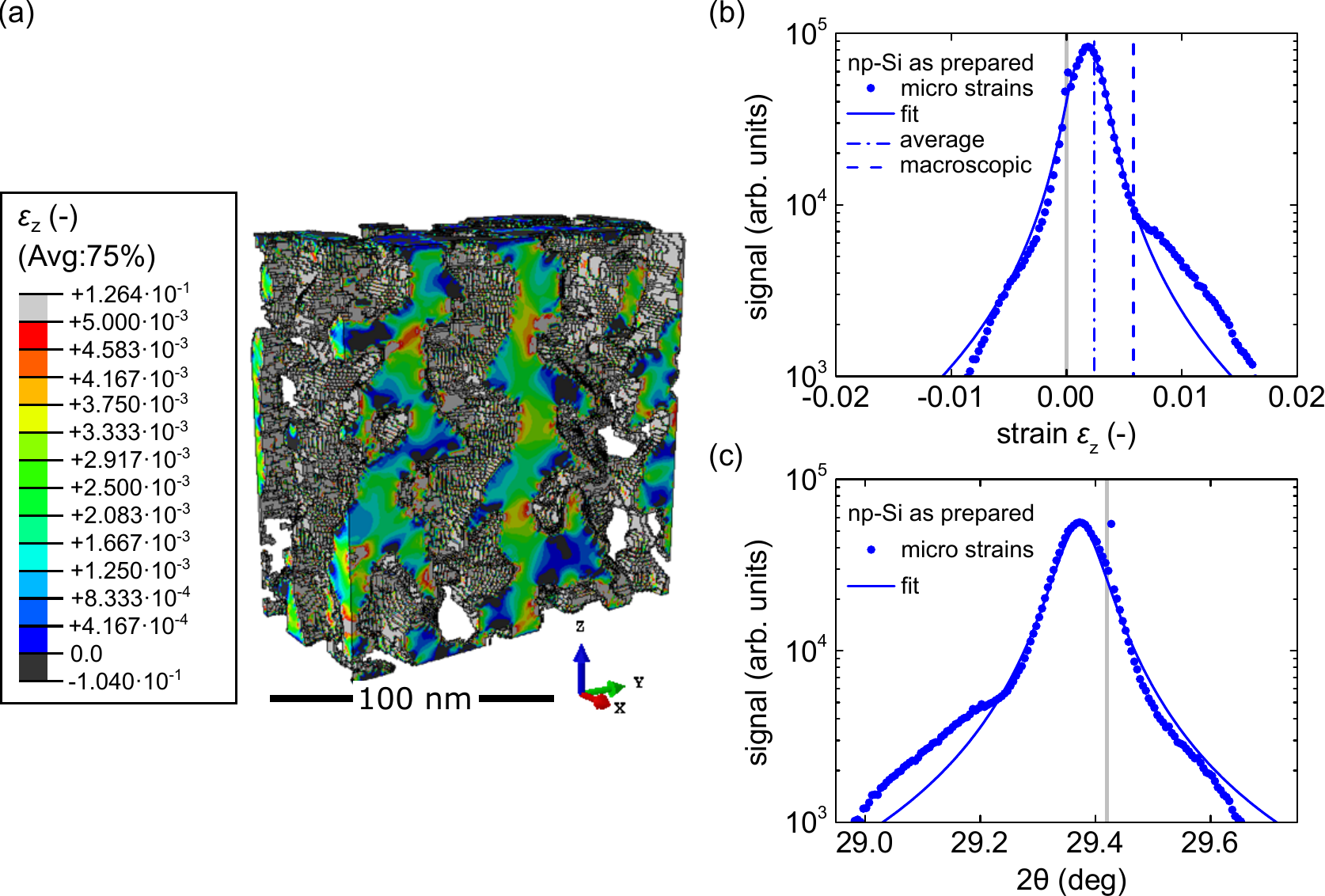}
		\caption{\textbf{Micromechanical model of the as-prepared np-Si/PPy hybrid.} (a) Distribution of micro strain $\varepsilon_z$ in nanoporous silicon strained by expansion of interface elements in the out-of-plane direction, i.e. $z$-direction in the model. The scale bar represents $100\,\mathrm{nm}$. (b) Histogram of strain $\varepsilon_z$ for all elements with peak position calibrated to the as-prepared state, where 'average' denotes the measured out-of-plane lattice mismatch of $\delta a/a_\perp=+(2.40\pm0.01)\cdot10^{-3}$. 'macroscopic' refers to the strain of the whole micromechanical model in comparison to the micro strains of single voxels. (c) Computed peak position by a transfer of micro strains in diffraction angle $2\theta$.}
		\label{Fig_epszSurf}
	\end{figure*}

\begin{figure}
		\centering
		\includegraphics{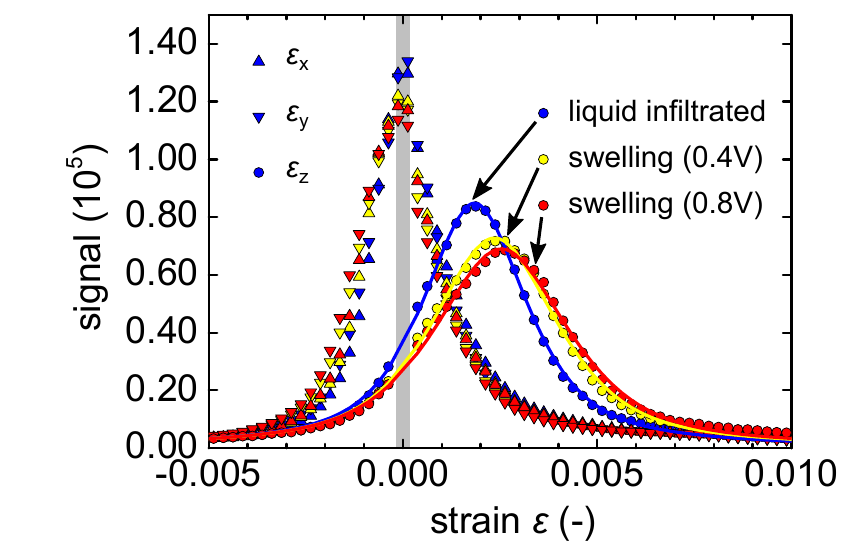}
		\caption{\textbf{Micromechanical simulation results.} Distribution of out-of-plane and in-plane strains in the RVE for different states: Histograms showing the increase of out-of-plane strain along np-Si with a PPy filling, liquid infiltrated and with applied potentials of $0.4\,\mathrm{V}$ and $0.8\,\mathrm{V}$ -- the assignment of symbols is given in the legend on the left, the different colors denote the different states.}
		\label{Fig_swellingSimulation1}
	\end{figure}

	%
	In the following a micromechanical analysis of the electrochemical actuation is presented, giving mechanistic insights on the experimental findings with regard to the strain states on the single-pore scale. Furthermore, the micromechanical simulation models the actuation strains on the microscopical level and assesses the macroscopic actuation of the sample upon electrolyte imbibition and applied potentials. 
	The materials and methods section gives a detailed description of the model's generation and its properties. The micromechanical simulation model is built from a TEM tomography measurement, conducted on the np-Si/PPy hybrid. A sample volume, extracted from the TEM tomography data, is converted into a finite element (FEM) model, by dividing it into small elements. The whole np-Si model is referred to as the representative volume element (RVE). Each small FEM element can be evaluated individually in terms of its mechanical properties. Thus, the whole model allows to study the strain distribution on the micro-structural length scale of the RVE.\\ 
	In the following, these resolved strains, on the level of a single element, are called micro strains. As the X-ray beam has a larger diameter than the size of our RVE, for a comparison with the experimental values, the micro strains in the out-of-plane direction, i.e. $z$-direction, are averaged from all elements of the model, that represent silicon. The model is calibrated, so that for the as-prepared and liquid infiltrated state, the lattice strains from the XRD measurements are reproduced, as described in the methods section. The results are shown in Figure \ref{Fig_epszSurf}. Fig. \ref{Fig_epszSurf}(a) depicts the micro strains, which occur in $z$-direction in the model corresponding to the out-of-plane direction of the np-Si/PPy hybrid. Figure \ref{Fig_epszSurf}(b) shows the quantitative analysis of the micro strains. The micro strains of the elements are depicted as dots and form a peak. The lattice mismatch in the out-of-plane direction, that was determined experimentally, is also depicted. For the as-prepared np-Si/PPy hybrid sample it is $\delta a/a_\perp=+2.40\cdot10^{-3}$, cf. \ref{Fig_XRD-out-ofplane}(c). In Fig. \ref{Fig_epszSurf}(b) it is denoted by a line as 'average'. It can be seen, that the experimental results are qualitatively reproduced by the resulting distribution of the micro strains, as the peak of the micro strain distribution matches well with the 'average' position. Figure \ref{Fig_epszSurf}(c) displays the angle $2\theta$, which would result from the micro strain distribution, if they are converted into a diffraction profile. Interestingly, the shape of the peak, qualitatively reproduces the shape of the measured np-Si/PPy peak. The peak from the simulation has a shoulder on its left side, towards smaller angles. The measured np-Si/PPy peak has a similar asymmetric distortion, as can be seen in Figure \ref{Fig_XRD-out-ofplane} (c). However, the width of the micro strain distribution derived from the simulation of np-Si in \ref{Fig_epszSurf}(c) is much wider compared to the signal in the experiment, which is shown in Fig. \ref{Fig_XRD-out-ofplane}(b) and (c). This is likely caused by the highly irregular microstructure, which leads to a large spread in the micro strain distribution and asymmetric peak shapes \cite{abramof2006}. The X-ray would average over several RVEs and thus irregularities would have a smaller impact. It should be stressed here, that for this precise reason, the RVE is representative for the mechanics of the material, but not its X-ray diffraction properties. Thus, the two graphs, from simulation and experiment, can not be compared directly in a quantitative manner. Therefore, solely a qualitative statement about the general shape of the peak can be made.\\
	The position denoted 'macroscopic' in Fig. \ref{Fig_epszSurf}(b) refers to the strain, that the whole micromechanical model, made of single voxel elements, exhibits. Therefore, not the strain of each element, i.e. the micro strains, is evaluated, but how the whole RVE model expands or contracts. Interestingly, the macroscopic strain in $z$-direction with a value of $\delta a/a_\mathrm{macroscopic}=+5.8\cdot10^{-2}$ is more than a factor of two larger than the peak of the micro strains in the RVE, which is adjusted to the measured out-of-plane lattice mismatch of $\delta a/a_\perp=+(2.40\pm0.01)\cdot10^{-3}$. This discrepancy between the microscopic and macroscopic level is apparently caused by the irregular pore structure, that, driven by the surface strain, expands omnidirectional. Due to the macroscopic clamping of the lateral displacements, this leads to an additional contribution in the macroscopic strain in out-of-plane, $z$-direction and therefore for an expansion of the whole RVE.\\
	The expansion caused by the potential driven incorporation of ions into PPy, is modeled with the thermal expansion of the PPy phase. Here, thermal expansion is only used to mimic the expansion of the PPy phase through the incorporation of ions. An ion-absorption-induced expansion can be reproduced well by a thermal expansion in an FEM model\cite{Brinker2020}. Again, the measured positions of the out-of-plane peaks, listed in Table \ref{Fig_Table}, have been used to calibrate the thermal expansion coefficient of the PPy phase via the averaged strains in $z$-direction of the silicon phase in the RVE. This results in a volumetric swelling of the PPy phase of $2.3\,\%$ and $3.2\,\%$ for the two applied potentials $0.4\,\mathrm{V}$ and $0.8\,\mathrm{V}$. Simulations for the different RVEs exhibit only a small variation in the predicted averaged strains. The different RVEs are shown in Fig. \ref{Fig_TEMstack} in the methods section. The as-prepared, liquid infiltrated state and the state with an applied potential of $0.8\,\mathrm{V}$ have standard deviations of $7.2\cdot10^{-5}$ and $9.3\cdot10^{-5}$, respectively. For PPy with applied potential of $0.8\,\mathrm{V}$, this scatter is approximately a factor of two larger than in the experiment. It can be explained by the small size of the RVE relative to the volume that is averaged by the synchrotron beam with varying microstructure and filling fractions.\\
	The predicted strain distributions in the silicon phase are shown in Fig. \ref{Fig_swellingSimulation1} for each direction of the simulation, i.e. the two in-plane directions $\varepsilon_\mathrm{x}$ and $\varepsilon_\mathrm{y}$, and the out-of-plane direction $\varepsilon_\mathrm{z}$. 
	\begin{figure*}
		\centering
		\includegraphics{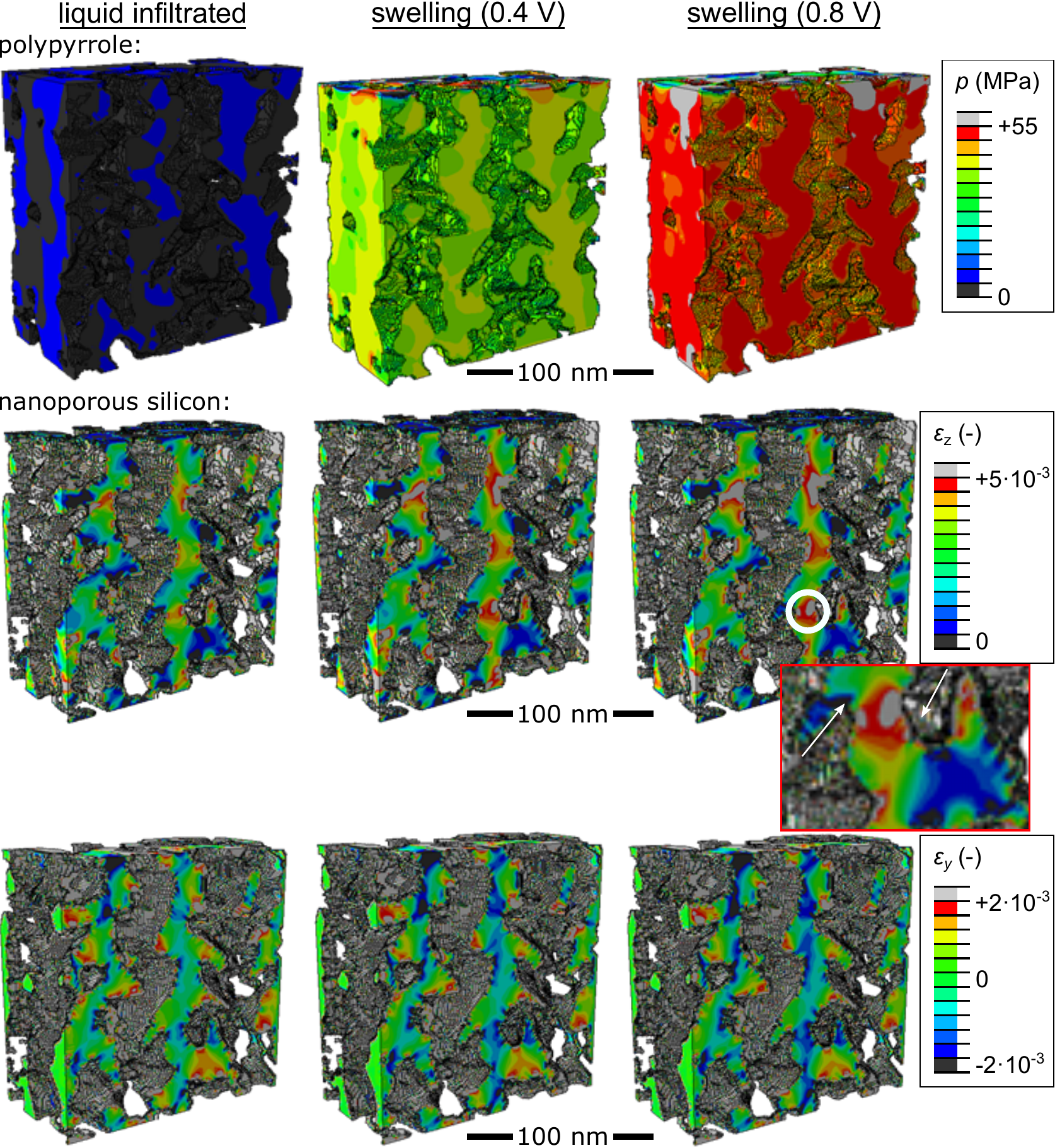}
		\caption{\textbf{Micromechanical simulation results with images of the representative volume element.} Pressure and strain distributions for the np-Si/PPy RVE for different states of liquid infiltrated and with applied potentials of $0.4\,\mathrm{V}$ and $0.8\,\mathrm{V}$. The upper row depicts the pressure inside the PPy phase. The middle and lower row show the silicon phase with strains in the out-of-plane, $z$-direction (middle row) and in the in-plane, $y$-direction (lower row). White circles mark the notch regions with high out-of-plane strains. The notch region is also shown in a close-up in the middle row with arrows marking the dendritic side pores. The scale bar represents $100\,\mathrm{nm}$.}
		\label{Fig_swellingSimulation2}
	\end{figure*}
	The strain distributions are furthermore depicted for the three states of liquid infiltrated PPy and with applied potentials of $0.4\,\mathrm{V}$ and $0.8\,\mathrm{V}$. The resulting pressure evolution in the PPy phase is presented in the upper row of Fig. \ref{Fig_swellingSimulation2}, ranging from zero in the liquid infiltrated state to $55\,\mathrm{MPa}$ at an applied potential of $0.8 \mathrm{V}$ for the PPy infiltrated material. Remarkably, this pressure level is about five times larger than the pressure that has been determined in another computational approach with a $2\,\sfrac{1}{2}\,$D model~\cite{Brinker2020} and this value does not substantially reduce by removing the clamping boundary conditions at $x=1$ and $y=1$. \\
	In Fig. \ref{Fig_swellingSimulation1}, the computed in-plane strains $\varepsilon_\mathrm{x}$ and $\varepsilon_\mathrm{y}$, i.e. the in-plane strains, show a small but noticeable shift to negative values with the higher potential applied. The out-of-plain strain $\varepsilon_\mathrm{z}$ shows a proportionality with a positive sign on the strain, i.e. the peak moves further away from zero with increasing potential. This is in very good agreement with the experimental observations, see Fig. \ref{Fig_XRD_in-plane}(d). The behavior can be explained by evaluating the strain distributions depicted in Fig. \ref{Fig_swellingSimulation2}. From the middle row of Fig. \ref{Fig_swellingSimulation2}, displaying the out-of-plane strain distribution $\varepsilon_z$ in the silicon phase, it can be seen that higher strains are located in the thin cross-sections that act like notches, while the thicker cross-sections remain almost unstrained. At these notches two dendritic side pores penetrate and restrain the pore wall. Below and above this notch it can be seen that the pore wall is thicker. These regions coincide with lower strains in the $z$-direction, whereas directly in the notch region higher strains occur.\\
	The in-plane strain component $\varepsilon_y$ is shown in the bottom row. It is visible from the scale of the strain, that negative $\varepsilon_y$ strains are present in the RVE, visible as dark blue color. The regions, where these negative strains occur, correlate with the notched and highly strained regions in $\varepsilon_z$ (middle row). Therefore, the negative shift of the peak for in-plane components, shown in Figure \ref{Fig_swellingSimulation1}, is a result of transverse contraction in regions that show large tensile strains in $z$-direction.\\ 
	%
	%
	%
	%
	%
	%
	%
	%
	%
	%
	\section*{Discussion}
	First, the $\theta-2\theta$ XRD measurement of the dry sample in the reflection geometry, as depicted in Figure \ref{Fig_XRD-out-ofplane}(b), is examined in detail. The reason the $\theta-2\theta$ measurement shows two distinctive peaks results from a different lattice parameter of np-Si to bulk silicon. Already solely due to the introduction of the pores into silicon, a distinct surface stress evolves at the interface of np-Si pore wall and the ambient pore space and forces that act on the interface lead to a distinct difference of $a_\mathrm{np-Si}$~\cite{Dolino1996}. np-Si's lattice constant can be further shifted by infiltrating the pores, e.g., by the adsorption of gas species~\cite{Dolino1996}, since this causes either adsorption-induced changes in the interfacial stress (Bangham effect) or capillary forces due to liquid condensation in the pores (Laplace pressure effects)~\cite{Grosman2015,Gor2015,Gor2017, Rolley2017, Gor2018}.\\
	The np-Si layer can be considered as clamped, as it is attached to a bulk silicon substrate. Therefore, the planes perpendicular to the interface are constrained to the interatomic spacing of the underlying bulk silicon. To compensate for the in-plane clamping, the lattice constant of the attached np-Si layer in out-of-plane direction is increased further~\cite{Barla1984}. Hence, an XRD measurement produces two distinctive peaks for both bulk- and np-Si in the out-of-plane direction. It should be noted, that the mismatch of lattice parameters also leads to a curvature of the sample. In the case of a plane, unfilled np-Si layer the curvature radius is found to not affect the XRD measurement in a significant way, when the diffraction area is limited~\cite{Barla1984}. Moreover, the thickness of the bulk silicon substrate in this work, amounting to roughly $500\,\mathrm{\upmu m}$, is chosen to further suppress bending. However, the deviation from an ideal Bragg peak shape of the bulk silicon peak at $2\theta=(29.4159\pm0.0003)\,^{\circ}$ is striking. It has apart from the main peak a broad shoulder to the smaller $2\theta$ values with an additional side peak at $2\theta=29.4094\,^{\circ}$. Inhomogeneities in the polymer filling of the pores lead to a different stress transfer onto the bulk silicon substrate and the accompanying curvature. Hence, this would yield a broader distribution of the lattice parameter and thus the bulk silicon Bragg peak. Furthermore, a broad foot towards higher values of $2\theta$ from $2\theta=29.4317\,^{\circ}$ on at the bottom of the main peak is visible. This feature might be a result of a diffuse scattering of the porous structure~\cite{Buttard1996}.\\
	A similar distortion is present on the np-Si peak at $2\theta=(29.3437\pm0.0003)\,^{\circ}$, as can be seen in Figure \ref{Fig_XRD-out-ofplane}(b). The peak deviates from a Bragg peak with a Gaussian or Lorentzian form and has a rather broad shape with an asymmetry towards the flank at smaller $2\theta$ values. The occurrence of this broadening of the peak might be due to an irregular local filling of the pores by PPy. The stress caused on the silicon polymer interface in the single pore then might be highly diverging according to the actual filling of the respective single pore. Moreover, an irregularity might lie in the morphology of the pores. Thereby, smaller pore sections branch off from the main pore in a non regular fashion~\cite{Sailor2011,Brinker2020}. Hence, these small side-pores might experience a wide filling distribution, which in turn would lead to an irregular stress distribution on the np-Si pore walls. A similar distortion has been reported for oxidized np-Si. The stresses caused by the different silicon-silicon atom spacing in the oxide compared to bulk silicon through a progressive oxidation of the pore walls leads to a straining of the lattice. Hence, the width of the Bragg peak increases and its peak intensity decreases, i.e. the peak is broadened~\cite{Buttard1996}.\\
	The results for the lattice mismatch obtained in the experiments are summarized in Fig. \ref{Fig_Table}. The lattice mismatch obtained from the non-liquid-infiltrated sample, discussed above and presented in Figure \ref{Fig_XRD-out-ofplane}(b), amounts to $\delta a/a_\perp=+(2.40\pm0.01)\cdot10^{-3}$, which means an expansion of the lattice in the out-of-plane direction. This characteristic is well known from literature~\cite{Barla1984,Dolino1996,Bellet1993, Bellet1994,Bellet1996,Buttard1996}. As-synthesized np-Si, prepared from silicon with an identical resistivity and a porosity equal to the here presented material, exhibits a lattice mismatch of $\delta a/a_\perp=+5.2\cdot10^{-4}$~\cite{Buttard1996}. Hence, the lattice expansion of pure np-Si is about an order of magnitude smaller than the here presented np-Si/PPy hybrid. Thus, the polymer filling alone strains the np-Si lattice significantly and expands the np-Si in the out-of-plane direction. For np-Si, prepared under different conditions, a lattice mismatch value as large as $\delta a/a_\perp=+0.002$ for plain np-Si has been reported before~\cite{Buttard1996}. The lattice mismatch is thereby in the same order of magnitude as the value for the hybrid showing that large out-of-plane strain as such is achievable, despite a vastly different pore morphology for the referenced np-Si material.\\
	The $\theta-2\theta$ scan conducted in transmission geometry for a sample in perchloric acid electrolyte solution, depicted in Figure \ref{Fig_XRD_in-plane}(b), reveals a second smaller peak towards higher $2\theta$. This additional peak is not present for an as-prepared np-Si layer, as can be seen when comparing the two signals in the figure. Such a distinct peak in the in-plane direction is not to be expected, since the bulk silicon substrate clamps the lattice in the np-Si layer and stress induced by the lattice mismatch releases via the out-of-plane lattice component~\cite{Barla1984}. Even the stress exerted through an oxidation of the np-Si does not lead to an additional peak in the in-plane direction, as an estimate gives $0\pm5\cdot10^{-5}$ for the in-plane direction~\cite{Buttard1996}. On the here reported sample the lattice mismatch amounts to $\delta a/a_{||}=-(1.17\pm0.07)\cdot10^{-4}$ and thus beyond the referenced margin of uncertainty. The observed second peak and the accompanying in-plane lattice mismatch is caused with certainty by the considerable stress of the PPy filling of the pores. Interestingly, as the np-Si peak is at larger $2\theta$ values, the lattice mismatch is negative. The lattice is contracted in the in-plane direction. Assuming a positive Poisson's ratio, an in-plane contraction is to be expected with an out-of-plane expansion. Indeed, the micromechanical results trace this in-plane contraction to regions, that are highly strained in the out-of-plane direction and, thus, to transverse contraction. In total, the measurements on the np-Si/PPy hybrid in the as-prepared state in both measurement geometries produce comprehensible results.\\
	Next, the effect of the imbibition of electrolyte solution into the PPy filled pores is further discussed. The effect on the lattice mismatch in the out-of-plane direction amounts to $\Delta (\delta a/a_\perp)=+(8\pm1)\cdot10^{-5}$. A similar experiment carried out on plane np-Si without a polymer filling examined the effect of pentane vapor adsorption on the out-of-plane lattice constant~\cite{Dolino1996}. Here, a difference of $\Delta (\delta a/a_\perp)=+1.8\cdot10^{-4}$ between the dry and the fully filled, liquid state was found. The effect has thus the same dependence with a positive sign, meaning that the adsorption of liquid leads to a further expansion of the np-Si lattice in the out-of-plane direction. This corresponds to the Bangham effect, i.e., the reduction of the surface stress upon molecular adsorption at the pore walls~\cite{Gor2017}. However, the effect for the pentane adsorption is about twice as large as the effect visible in our np-Si/PPy hybrid for perchloric acid. One reason may be due to different sample properties. The effect of pentane adsorption is measured in np-Si with a porosity of $80\,\%$, which is almost twice as large as the here determined porosity. Therefore, surface stress changes have a larger effect as np-Si becomes less stiff with increasing porosity~\cite{Barla1984}. PPy inside the pore does not constitute a complete filling. Rather, the polymer within the pores is porous itself and has a lower density than its bulk counterpart~\cite{Schultze1995,Brinker2020}. Thus, the imbibition of the liquid potentially leads to a compression and thereby a higher density of the polymer.\\
	As a next point the potential dependence of both lattice mismatches, out-of-plane and in-plane, are discussed. Increasing the potential leads to an increased incorporation of anion charge carriers and a reduction of the potential results in an expulsion of those charge carriers. The effect of the potential on the out-of-plane lattice mismatch is reversible and linear, as the stepwise potential- and the CV method reveal, cf. Figure \ref{Fig_XRD-out-ofplane}(e) and Figure \ref{Fig_time}(a). The amplitude of the lattice actuation is $A_{\delta a/a_\perp}=(1.50\pm0.04)\cdot10^{-4}$ and is about twice as large as the effect of the expansion through imbibition of electrolyte solution. The amplitude in the in-plane direction amounts to $A_{\delta a/a_{||}}=(1.06\pm0.09)\cdot10^{-5}$, which renders the in-plane effect of the applied potential on the actuation about an order of magnitude smaller.\\
	The kinetics of the electrochemical actuation on the out-of-plane lattice constant are measured and result in time scales for the reaction between $1$ and $3\,\mathrm{s}$ for the lattice mismatch and $7$ to $8\,\mathrm{s}$ for the charge movement. A kinetics measurement on the macroscopic scale leads to equal results~\cite{Brinker2020}. Thereby, the strain evolution of a fully porous membrane exposed to a square potential from $0.4-0.8\,\mathrm{V}$ is measured. The strain response with $2.8\,\mathrm{s}$ and $4.3\,\mathrm{s}$ is in the same range as the response of the lattice mismatch, whereas the charge carrier incorporation and expulsion exhibits characteristic times of $16.52\,\mathrm{s}$ and $14.31\,\mathrm{s}$~\cite{Brinker2020}. The slightly higher referenced values of $\tau$ with respect to both lattice mismatch and the charging might be due to a thicker np-Si material of $85\,\mathrm{\upmu m}$. In total these results measured at an macroscopic sample fit excellently to the here presented kinetic measurements. The reason for the significant difference between the strain response and the ion movement kinetics is a diffusion limitation, which hinders a faster charge carrier transport to counterbalance the abrupt potential change, see also the section C in the methods and materials. Generally, the manner in which the potential is applied has a significant impact on the amount and time-scale of charge incorporated in a porous material such as np-Si/PPy\cite{Breitsprecher2020}. Therefore, a diffusion limitation would influence the actuation response. Additionally, PPy possibly reaches its yield limit. Thereby, PPy starts to deform plastically and thus it is not straining the lattice any more while still incorporating charge carriers~\cite{Brinker2020}.\\
	A measurement of the macroscopic in-plane strain response to an applied potential on the np-Si/PPy hybrid in the same potential range, i.e., $0.4-0.9\,\mathrm{V}$, reveals a strain amplitude of $(5.0\pm0.2)\cdot10^{-4}$~\cite{Brinker2020}. The strain response is about 37 times larger than the in-plane lattice strain response, here presented. Moreover, the potential dependence is of proportionality with an opposite sign. Therefore, by increasing the potential the whole macroscopic hybrid expands and it contracts upon a reduction of the potential due to swelling and contraction of the incorporated PPy. To stress this point, here the strain is experimentally determined only on the scale of the np-Si lattice. In the actuation process, PPy builds up pressures of up to $15\,\mathrm{MPa}$ and causes stresses of $50$ to $100\,\mathrm{MPa}$ in the np-Si network~\cite{Brinker2020}. The pressure the PPy exerts on the np-Si walls then would be responsible for the here observed in-plane lattice compression, while a macroscopic expansion of the sample is observed.\\
	The micromechanical simulations based on TEM tomography data help to bring these results on the macroscopic strain response of the entire hybrid and the np-Si lattice response on the microscopic level in accordance to each other. 
	First of all, the predicted Young's modulus is anisotropic, i.e., the out-of-plane stiffness is much lower as it is expected for the common assumption of a pore structure that is oriented out-of-plane~\cite{Thelen2021}. Therefore, a meaningful interpretation is only possible with micromechanical simulations that make use of the full 3D information.\\
	First of all, a remarkable result is that the pressure in the PPy phase at an applied potential of $0.8\,\mathrm{V}$ is about five times larger than the pressure that has been determined with a $2\,\sfrac{1}{2}\,$D model~\cite{Brinker2020}. This is caused by the clamping of the np-Si layer on the bulk silicon substrate in the current setup, i.e., in contrast to~\cite{Brinker2020}, the RVE is constrained in lateral direction and is hindered to expand.\\
	Under this boundary conditions, both the surface strain of np-Si and the swelling of PPy result in a macroscopic expansion that is directed out of plane. This leads to stress concentrations in thin cross-sections of the vertical silicon pore walls and, consequently, to concentrations of negative transverse strains in those 'notched' regions. The remaining thicker cross-sections of the silicon matrix remain almost unstrained, which explains why the measured shift of the peak in the transmission geometry is a magnitude smaller than in the reflection geometry. The effect of the external pressure on the silicon walls in the in-plane direction is comparably small. Otherwise the strain distribution should correlate with the pressure in the PPy phase, which is much more homogeneously distributed. Finally, the micromechanical simulations show that the shift of the peaks correlates with the applied potential. The analysis of the actuation kinetics that is reflected in the measured time constants in the reflection geometry indicates that the PPy phase undergoes inelastic deformations that are not captured by the micromechanical simulation. This is in agreement with the findings by Brinker et al.~\cite{Brinker2020}, where also plastic deformations in the PPy phase are reported.\\ 
	\section*{Summary}
	\noindent We present high-resolution in operando X-ray diffraction experiments on the electrochemical actuation of a np-Si/PPy hybrid in an aqueous electrolyte.
	The information on the electrostrains in the np-Si pore walls could be modeled by micromechanical analysis based on full 3-D reconstructions of a representative volume of the nanoporous medium.\\
	Overall, the simulation successfully linked the microscopic to the macroscopic level and indicates that the microstructure has significant influence on the micro strains and consequently also on the effective electroactuorics.\\
	The in-plane mechanical response is dominantly isotropic despite the anisotropic elasticity of crystalline silicon, which is in agreement with recent laser-ultrasound experiments on similarly fabricated np-Si \cite{Thelen2021}. Presumably this is caused by the high porosity along with the irregular pore shapes.\\
	The pore structure anisotropy originating from the parallel alignment of the nanopores lead to significant differences between the in- and out-of-plane electromechanical response. Furthermore, the simulations highlight that the dendritic shape of the silicon pore walls, including pore connections between the main channels, cause complex, highly inhomogeneous stress-strain fields in the crystalline host on a microscopic level that result in the observed macroscopic, anisotropic strain response.\\
	Recent experiments on self-diffusion~\cite{Kondrashova2017} and cavitation events~\cite{Bossert2021} in this porous medium hinted towards the importance of those pore interconnections and questioned the simple picture of nanoporous silicon as an array of parallel independent nanochannels. Also our study evidences that a consideration of nanoporous silicon as a parallel array of independent pores on a random hexagonal array is insufficient.\\
	Time-dependent X-ray scattering experiments on the dynamics of the actuorics hint towards the importance of diffusion limitations, plastic deformation and creep in the nanoconfined polymer upon (counter-) ion ad- and desorption, the very pore-scale processes causing the macroscopic electroactuation. The slow ion diffusion results from the strong confinement effects. Here an hierarchical structuring of the porous host could lead to an improved performance \cite{Gries2019,Gries2022}.\\
	From a more general perspective, our study is a fine example that the combination of TEM tomography-based micromechanical simulations with high-resolution X-ray scattering experiments provides a powerful approach for in operando analysis of nanoporous composites from the single-nanopore up to the porous-medium scale.\\
	Finally, we envision analogous combinations of in operando X-ray diffraction and tomography experiments as a function of pore morphology, specifically for nanoporous silicon synthesized from $($111$)$ wafers, where indeed a full 3-D isotropic behavior is expected~\cite{Vincent2017}.  
	\medskip
	\begin{acknowledgements} \par 
		\noindent This work was supported by the Deutsche Forschungsgemeinschaft (DFG) within the Collaborative Research Initiative CRC 986 "Tailor-Made Multi-Scale Materials Systems" Project number 192346071. This project has also received funding from the European Innovation Council (EIC) under the European Union's Horizon 2020 research and innovation program under grant agreement No. 964524 EHAWEDRY: "Energy harvesting via wetting-drying cycles with nanoporous electrodes" (H2020-FETOPEN-1-2021-2025). We thank Deutsches Elektronen-Synchrotron DESY, Hamburg for access to the beamline P08 of the PETRA III synchrotron. M.M. acknowledges support by the Deutsche Forschungsgemeinschaft (DFG) within the project "Hybrid Thermoelectric Materials Based on Porous Silicon: Linking Macroscopic Transport Phenomena to Microscopic Structure and Elementary Excitations" project number 402553194. We thank Tommy Hofmann (Helmholtz-Zentrum Berlin für Materialien und Energie) for helpful discussions and a critical reading of the manuscript. We also acknowledge the scientific exchange and support of the Centre for Molecular Water Science CMWS, Hamburg (Germany).\\
		\textbf{Data and materials availability} \par
		\noindent All data that is needed to evaluate the conclusions is presented in the paper and in the supplementary materials. The electrochemical and x-ray raw data of the electrochemical actuation experiments as well as the reconstructed TEM-based tomography model are available at TORE (https://tore.tuhh.de/), the Open Research Repository of Hamburg University of Technology, at the doi: https://doi.org/10.15480/336.4675.\\
		\textbf{Author contributions} \par 
		\noindent M.B., M.T. and P.H. conceived the experiments. M.B. performed the material synthesis. M.B., M.T., M.M., P.L., T.F.K. and P.H. performed the X-ray scattering experiments at the synchrotron PETRA III at Deutsches Elektronen-Synchrotron DESY. F.B. participated in the X-ray experiments as the beamline scientist. The data analysis was performed by M.B., M.T., P.L., F.B. and P.H. The TEM tomography and the respective sample preparation was carried out by D.R. and T.K. N.H. performed the micromechanical modeling and the analysis of the modeling data. M.B., N.H. and P.H. wrote the manuscript. All authors revised the manuscript.\\ 
		The authors declare that they have no competing interests.\\
	\end{acknowledgements}
	%
	%
	%
	\medskip
	\clearpage
	%
\end{document}